\pdfoutput=1
\documentclass[12pt, draftclsnofoot, onecolumn]{IEEEtran}
\usepackage[noend]{algorithmic}
\usepackage{algorithm}
\usepackage{amsmath}
\usepackage{amsfonts}
\usepackage{amssymb}
\usepackage{bm}
\usepackage{graphicx}
\usepackage{color}
\usepackage{multirow}
\usepackage{tabularx}
\usepackage[noend]{algorithmic}
\usepackage{algorithm}
\usepackage{epstopdf}
\newtheorem{lem}{Lemma}
\newtheorem{theorem}{{Theorem}}
\newtheorem{prob}{Problem}
\newcommand{\PreserveBackslash}[1]{\let \temp =\\#1 \let \\ = \temp}
\newcolumntype{C}[1]{>{\PreserveBackslash\centering}p{#1}}
\newcolumntype{R}[1]{>{\PreserveBackslash\raggedleft}p{#1}}
\newcolumntype{L}[1]{>{\PreserveBackslash\raggedright}p{#1}}

\begin{document}
\title{Dynamic Spectrum Leasing with Two Sellers}

\author{Rongfei Fan, \IEEEmembership{Member,~IEEE,} Wen Chen, Hai Jiang, \IEEEmembership{Senior Member,~IEEE}, Jianping An, \IEEEmembership{Member,~IEEE,} Kai Yang, \IEEEmembership{Member,~IEEE,} and Chengwen Xing, \IEEEmembership{Member,~IEEE} 
\IEEEcompsocitemizethanks{\IEEEcompsocthanksitem R. Fan, W. Chen, J. An, K. Yang, and C. Xing are with the School of Information and Electronics, Beijing Institute of Technology, Beijing, 100081, P. R. China. E-mail: \{fanrongfei,chenwen93,an\}@bit.edu.cn, yangkbit@gmail.com,  xingchengwen@gmail.com.\protect\\
H. Jiang is with the Department of Electrical and Computer Engineering, University of Alberta, Edmonton, AB T6G 1H9, Canada E-mail: hai1@ualberta.ca.
}
}

\IEEEtitleabstractindextext{%
\begin{abstract}
This paper studies dynamic spectrum leasing in a cognitive radio network. There are two spectrum sellers, who are two primary networks, each with an amount of licensed spectrum bandwidth. When a seller has some unused spectrum, it would like to lease the unused spectrum to secondary users. A coordinator helps to perform the spectrum leasing stage-by-stage. As the two sellers may have different leasing period, there are three epochs, in which seller 1 has spectrum to lease in Epochs II and III, while seller 2 has spectrum to lease in Epochs I and II. Each seller needs to decide how much spectrum it should lease to secondary users in each stage of its leasing period, with a target at revenue maximization. It is shown that, when the two sellers both have spectrum to lease (i.e., in Epoch II), the spectrum leasing can be formulated as a non-cooperative game. Nash equilibria of the game are found in closed form. Solutions of the two users in the three epochs are derived.
\end{abstract}

}

\maketitle

\IEEEdisplaynontitleabstractindextext
\IEEEpeerreviewmaketitle

{\section{Introduction}\label{sec:introduction}}

Cognitive radio has been considered as a promising solution to the spectrum shortage problem in the near future. In cognitive radio, if a primary user (a licensed user with some licensed spectrum bandwidth) has some unused spectrum for a certain amount of time, it may lease the unused spectrum to secondary users. By this method, the spectrum opportunities are exploited, and the primary user can earn extra payment from secondary users.

Spectrum leasing has been well investigated in the literature, in the modes of monopoly spectrum leasing (in which there is one spectrum seller) and oligopoly spectrum leasing (in which multiple spectrum sellers exist). In either mode, the research focus is on how to set the spectrum price.

In monopoly spectrum leasing, the major target is to achieve the maximal revenue of the seller. In the work of  \cite{Huangjianwei_2}, there are a spectrum provider, a broker, and a number of secondary users. By a Stackelberg game modeling, the broker optimally decides on the number of channels it should purchase from the spectrum provider as well as the price it should use to sell the purchased spectrum to secondary users. The work in \cite{Qianliang_monopoly} also considers a broker. It is assumed that for a given spectrum price, the amount of spectrum demand from secondary users is random. The work in \cite{Ashraf_monopoly} considers the impact of spectrum leasing on primary user performance (such as possible extra interference to the primary system). An optimal solution is given for the primary user, which strikes a balance between the earned revenue and the cost.

In oligopoly spectrum leasing, the major target is to achieve an equilibrium in the competition among multiple spectrum sellers.
Two brokers are assumed in \cite{Duanlingjie_duopoly}. Each broker decides on the amount of spectrum that it should purchase from spectrum providers and on the spectrum price that it should announce to secondary users, with a target at profit maximization. The work in \cite{Alexander_duopoly} also considers two brokers, by assuming that the leased spectrum may be shared by multiple secondary users simultaneously. Therefore, interference among secondary users needs to be taken into account.
The works in \cite{Zhu_1,Zhu_2,Lotfi} consider a duopoly market, in which the price competition of two spectrum sellers is investigated by using game theoretical approaches.
The work in \cite{Niyato_2} discusses the case with multiple sellers. By using an evolutionary game model, a solution is given for secondary users for their spectrum selection and for sellers for price setting.
The work in \cite{Jiajuncheng_multiple} considers multiple sellers as well as one broker, in which the impact of spectrum leasing on sellers' performance (i.e., service quality degradation) is taken into account.
The work in \cite{Gaurav_multiple} considers heterogenous secondary users, i.e., different secondary users may have different criteria on their spectrum leasing decisions.

In all above works, spectrum leasing is performed only once, and the price is fixed for the whole spectrum leasing duration, referred to as {\it static spectrum leasing}. On the other hand, {\it dynamic spectrum leasing}, in which the spectrum price may change over time, is more appropriate for the cases that the secondary users may need spectrum at different time instants.
In \cite{Huangjianwei_Lishuqin}, dynamic pricing in monopoly spectrum leasing is performed over infinite time horizon. The spectrum price is set dynamically, with a target of long-term average revenue maximization.
In \cite{Fan_Dynamic}, dynamic pricing in monopoly spectrum leasing is performed over a finite duration. The finite duration is divided into a number of stages, and the price in each stage is set up so as to maximize the overall revenue. To the best of our knowledge, there is no research in the literature on dynamic pricing with more than one spectrum seller.

\begin{figure*}
\begin{center}
\includegraphics[angle=0,width=4.5in]{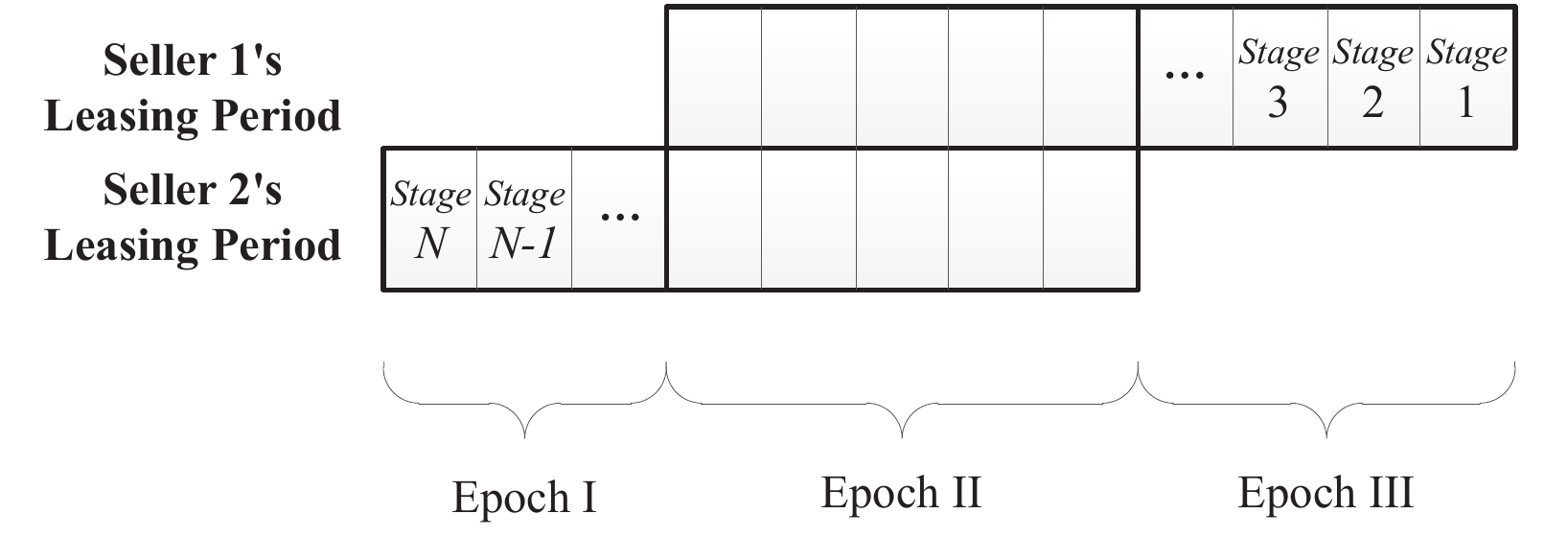}
\caption{Leasing periods of the two sellers. \label{fig_model}}
\end{center}
\end{figure*}

To fill the research gap, in this paper, we study dynamic spectrum leasing problem in a duopoly market with two sellers\footnote{We consider two sellers (i.e., a duopoly spectrum market) for the following reasons. 1) A duopoly spectrum market is a typical and popular scenario for cognitive radio, and has been adopted by many research efforts in the literature \cite{Duanlingjie_duopoly,Alexander_duopoly,Zhu_1,Zhu_2,Lotfi}. 2) Sufficient insights can be provided by the duopoly scenario into the spectrum leasing, and our method in this paper can be extended to the scenarios with more spectrum sellers, with increased complexity in analysis and presentation. For ease of analysis and presentation, we consider a duopoly scenario.}. As the two sellers may have different leasing periods, the system has three epochs, in which seller 1 has spectrum to lease in Epochs II and III, while seller 2 has spectrum to lease in Epochs I and II. The main contributions in this paper are summarized as follows. 1) We show that, the spectrum leasing problems of the sellers in Epoch I and Epoch III are convex optimization problems. For Epoch II, we formulate spectrum leasing of the two sellers as a non-cooperative game. We derive closed-form expressions for the Nash equilibria of the non-cooperative game. 2) The amount of spectrum that seller 1 would like to lease in Epoch III affects the non-cooperative game in Epoch II, and thus, affects the total revenues of the two sellers. By analyzing properties of seller 1's revenue in Epoch II and Epoch III, we propose a method that finds the optimal amount of spectrum that seller 1 should lease to secondary users in Epoch III.

The rest of this paper is organized as follows.
In Section \ref{s:system_model}, the system model is presented, and the spectrum leasing problems for the two sellers are formulated. In Section \ref{s:competetion}, Nash equilibria of the non-cooperative game in Epoch II are derived. Section \ref{s:monopoly} discusses how seller 1 should distribute its spectrum to be leased in Epoch II and Epoch III. Numerical results are given in Section \ref{s:numerical results}, and finally the paper is concluded in Section \ref{s:conclusion}.

\section{System Model and Problem Formulation} \label{s:system_model}
Consider two spectrum sellers (seller 1 and seller 2), one coordinator, and multiple secondary users. The coordinator is to help the two sellers to lease spectrum to secondary users. The two sellers are primary networks with a certain amount of licensed spectrum bandwidth. For each seller, when the data traffic from its own users is light, the seller may partition its spectrum bandwidth into two portions: primary portion and secondary portion. The primary portion will be assigned to the seller's own users, and the secondary portion can be leased to secondary users. In specific, consider that seller 1 and seller 2 have bandwidth $Q_1$ and $Q_2$ in their secondary portion, respectively. For each seller, the bandwidth in its secondary portion can be leased to secondary users for a duration (called {\it leasing period}). Consider that the two sellers' leasing periods are not identical,\footnote{If the two leasing periods are identical, it is a special case of the problem considered in this paper.} and overlap with each other.
Without loss of generality, we assume that the leasing period of seller 2 starts earlier than the leasing period of seller 1. We also assume that the leasing period of seller 2 ends earlier than that of seller 1.\footnote{Note that the method in this paper can be straightforwardly extended to deal with the case when the leasing period of seller 2 ends later than that of seller 1.} An illustration of the two leasing periods is given in Fig.~\ref{fig_model}. Here the union of the two leasing periods contains $N$ fixed-length stages. For presentation simplicity, the last stage of seller 1's leasing period is called stage 1, while the first stage of seller 2's leasing period is called stage $N$. Seller $i$ ($i=1,2$) would distribute its spectrum bandwidth $Q_i$ to be leased in the stages of its leasing period. In other words, it needs to decide on the amount of spectrum bandwidth to be leased in each stage in its leasing period, with a constraint that the total amount of leased spectrum bandwidth in the stages is bounded by $Q_i$. For seller $i$, denote the amount of spectrum bandwidth it would like to lease to secondary users in stage $n$ as $d_{n,i}$. At the beginning of stage $n$, seller $i$ should report to the coordinator the information of $d_{n,i}$.

At the beginning of stage $n$, after the coordinator gets the information of $d_{n,1}$ and $d_{n,2}$, it would set up a spectrum unit price (the price per unit bandwidth per stage) and lease the spectrum bandwidth $(d_{n,1}+d_{n,2})$ to secondary users. In other words, the coordinator should set up the unit price to attract $(d_{n,1}+d_{n,2})$ spectrum bandwidth demand from secondary users.
Denote the price $p$ to attract $d$ spectrum bandwidth demand as $P(d)$, which is a function of $d$. Economics analysis \cite{Ahn,Omar} has shown that price and demand typically follow a linear model, and thus, price $p$ and spectrum bandwidth demand $d$ satisfy the following feature:
\begin{equation} \label{e:price_def}
p=P\left(d\right)=C_0 - C_1 \cdot d
\end{equation}
in which $C_0$ and $C_1$ are coefficients.
$P(d)$ is a decreasing function of $d$. In addition, $d\cdot P(d)$ should be an increasing function of $d$ (as the total revenue for more leased spectrum bandwidth should be higher), based on which we have
\begin{equation} \label{e:C_0_C_1_assum}
C_0 > 2 C_1 \left(Q_1+Q_2\right).
\end{equation}


From Fig.~\ref{fig_model}, the union of the two sellers' leasing periods can be divided into three epochs: In Epoch I, only seller 2 has spectrum to lease; in Epoch II, both sellers have spectrum to lease; and in Epoch III, only seller 1 has spectrum to lease. Denote the set of stages in Epoch I, II, and III as $\mathcal{N}_{\text{I}}$, $\mathcal{N}_{\text{II}}$, and $\mathcal{N}_{\text{III}}$, respectively. Denote the set of stages in the leasing period of seller 1 and seller 2 as $\mathcal{N}_1$ and $\mathcal{N}_2$, respectively. Thus, we have $\mathcal{N}_1 = \mathcal{N}_{\text{II}}  \cup \mathcal{N}_{\text{III}}$ and $\mathcal{N}_2 = \mathcal{N}_{\text{I}}  \cup \mathcal{N}_{\text{II}}$.

Seller $i$ $(i\in \{1,2\})$ aims at maximizing its total revenue over all the stages by deciding on $d_{n,i}, n\in \mathcal{N}_i$. Next, the spectrum leasing problem in each epoch is discussed.

\subsection{Spectrum Leasing Problem in Epoch I}

In Epoch I, only seller 2 has spectrum to lease, and it does not know when seller 1 will join the spectrum leasing market and how much spectrum bandwidth seller 1 will offer for spectrum leasing. So seller 2 assumes a monopoly market. At a stage in Epoch I, once an amount of spectrum is leased to secondary users, the spectrum can be used by secondary users until the last stage of seller 2's leasing period.
Seller 2's collected revenue at stage $n$ is $ \left(C_0-C_1 d_{n,2}\right)  d_{n,2} \left(n-|\mathcal{N}_{\text{III}}| \right)$, in which $|\cdot |$ means cardinality of a set. To maximize its overall revenue, primary use 2 should solve the following optimization problem\footnote{In Epoch I, seller 2 does not know the value of $|\mathcal{N}_{\text{III}}|$. However, it knows the value of $\left(n-|\mathcal{N}_{\text{III}}| \right)$ (the length from stage $n$ until the end of seller 2's leasing period). Thus, in Problem \ref{p:Epoch_I}, we use notation $\left(n-|\mathcal{N}_{\text{III}}| \right)$, for consistence of the formulated spectrum leasing problems in the three epochs.}:
\begin{prob}\label{p:Epoch_I}
\begin{equation}
\begin{array}{cll}
\mathop {\max} \limits_{\{d_{n, 2}|n \in \mathcal{N}_{2}\}} & \sum\limits_{n \in \mathcal{N}_{2}} { \left(C_0-C_1 d_{n,2}\right)  d_{n,2} \left(n-|\mathcal{N}_{\text{III}}| \right)} \\
\text{s.t.} & \sum \limits_{n \in \mathcal{N}_{2}} d_{n,2} \leq Q_{2} \\
& d_{n,2} \geq 0, \forall n \in \mathcal{N}_{2}.
\end{array}
\end{equation}
\end{prob}
Problem \ref{p:Epoch_I} is a convex optimization problem. Thus the global optimal solution of Problem \ref{p:Epoch_I} can be achieved by existing numerical optimization methods.

\subsection{Spectrum leasing Problem in Epoch II}
At Epoch II's first stage (denoted as stage $l$), seller 1 has available spectrum bandwidth $Q_1$, while we denote the remaining spectrum bandwidth of seller 2 as $Q_2^\text{II}$ (in other words, spectrum bandwidth with amount $(Q_2 - Q_2^\text{II})$ has been leased out by seller 2 in Epoch I). At the beginning of stage $l$, each seller does not know the presence of the other seller, and thus, assumes a monopoly spectrum leasing. So each seller reports to the coordinator the amount of spectrum bandwidth it would like to lease to secondary users in the stage. In specific, seller 1 first solves the following convex optimization problem:
\begin{equation}\label{e:epochII_p1}
\begin{array}{cll}
\mathop {\max} \limits_{\{d_{n, 1}|n \in \mathcal{N}_{1}\}} & \sum\limits_{n \in \mathcal{N}_{1}} { \left(C_0-C_1 d_{n,1}\right)  d_{n,1} n} \\
\text{s.t.} & \sum \limits_{n \in \mathcal{N}_{1}} d_{n,1} \leq Q_{1} \\
& d_{n,1} \geq 0, \forall n \in \mathcal{N}_{1}
\end{array}
\end{equation}
and reports to the coordinator the values of  $d_{l,1}$ ($d_{l,1}$ is from the optimal solution of the above problem) and $|\mathcal{N}_1|$ (the leasing duration for the $d_{l,1}$ spectrum bandwidth). On the other hand, seller 2 reports to the coordinator the values of $d_{l,2}$ and ($|\mathcal{N}_2|-|\mathcal{N}_\text{I}|$) (which is the length of seller 2's remaining leasing period), in which $d_{l,2}$ is from the optimal solution of the following convex optimization problem:
\begin{equation}
\begin{array}{cll}
\mathop {\max} \limits_{\{d_{n, 2}|n \in \mathcal{N}_{2}\backslash \mathcal{N}_\text{I} \}} & \sum\limits_{n \in \mathcal{N}_{2}\backslash \mathcal{N}_\text{I}} { \left(C_0-C_1 d_{n,2}\right)  d_{n,2} \left(n-|\mathcal{N}_{\text{III}}| \right)} \\
\text{s.t.} & \sum \limits_{n \in \mathcal{N}_{2}\backslash \mathcal{N}_\text{I}} d_{n,2} \leq Q_{2}^\text{II} \\
& d_{n,2} \geq 0, \forall n \in \mathcal{N}_{2}\backslash \mathcal{N}_\text{I}.
\end{array}
\end{equation}

Then the coordinator feeds back to the two sellers by telling 1) that now two sellers have spectrum to lease, 2) how much spectrum bandwidth each seller offers in this stage, and 3) how long the leasing period is for each seller. From $d_{l,1}$ and $|\mathcal{N}_1|$ in the feedback information, seller 2 can find out the available stock of seller 1, by searching the value of $Q_1$ (using bisection search) that makes $d_{l,1}$ be in the optimal solution of the problem in (\ref{e:epochII_p1}). Similarly, seller 1 can also find out the available stock of seller 2. Based on stock information of the other seller, each seller adjusts the amount of offered spectrum bandwidth ($d_{l,1}$ or $d_{l,2}$) and resubmits to the coordinator, and the coordinator decides on a unit price based on (\ref{e:price_def}) with total spectrum demand $(d_{l,1}+d_{l,2})$. In each subsequent stage (say stage $n$) in Epoch II, by knowing the existence of the other seller, each seller reports to the coordinator the amount of offered spectrum bandwidth ($d_{n,1}$ and $d_{n,2}$), and the coordinator decides on a unit price based on (\ref{e:price_def}) with total spectrum demand $(d_{n,1}+d_{n,2})$. 

 In every stage in Epoch II, once an amount of spectrum bandwidth of a seller is leased to secondary users, the spectrum can be used by secondary users until the last stage of the corresponding seller's leasing period.

A decision that seller 1 should make in Epoch II is the amount $Q_1^\text{III}$ of spectrum bandwidth it reserves for Epoch III, where $Q_1^\text{III} \in [0, Q_1]$. In other words, seller 1 would like to lease spectrum bandwidth ($Q_1-Q_{1}^{\text{III}}$) in Epoch II.

In Epoch II, the announced unit price at each stage (say stage $n$) depends on the sum of $d_{n,1}$ and $d_{n,2}$. Thus, there is a non-cooperative game between the two sellers. In this game,
the two players are seller 1 and seller 2,
and the strategy of seller 1 and seller 2 are $\mathcal{S}_{1} \triangleq \{ d_{n, 1}|n\in \mathcal{N}_{\text{II}}\}$ and $\mathcal{S}_{2} \triangleq \{d_{n,2}|n\in \mathcal{N}_{\text{II}}\}$, respectively. The payoff function of seller 1 and seller 2 can be expressed as
\begin{equation*}
{R}_{1} \left(\mathcal{S}_{1}, \mathcal{S}_{2} \right) \triangleq \sum\limits_{n \in \mathcal{N}_{\text{II}}} { \left(C_0-C_1 \left( d_{n,1} + d_{n,2}\right)\right)  d_{n,1} n}
\end{equation*}
and
\begin{equation*}
{R}_{2} \left(\mathcal{S}_{1}, \mathcal{S}_{2} \right) \triangleq \sum\limits_{n \in \mathcal{N}_{\text{II}}} { \left(C_0-C_1\left(d_{n,1} + d_{n,2}\right)\right)  d_{n,2} \left(n-|\mathcal{N}_{\text{III}}|\right)},
\end{equation*}
respectively.
Define the feasible region of seller 1's strategy as
\begin{equation*}
\mathcal{F}_{1}\left(y\right) = \left\{\{d_{n,1}|n\in \mathcal{N}_{\text{II}} \} \Big|\sum \limits_{n \in \mathcal{N}_{\text{II}}} d_{n,1} \leq y,  d_{n,1} \geq 0\right\}
\end{equation*}
when seller 1 would like to lease to secondary users spectrum bandwidth amount $y$ in Epoch II,
and define the feasible region of seller 2's strategy as
\begin{equation*}
\mathcal{F}_{2}\left(z\right) = \left\{\{d_{n,2} |n\in \mathcal{N}_{\text{II}} \} \Big|\sum \limits_{n \in \mathcal{N}_{\text{II}}} d_{n,2} \leq z,  d_{n,2} \geq 0\right\}
\end{equation*}
when seller 2 would like to lease to secondary users spectrum bandwidth amount $z$ in Epoch II.
The objective of seller 1 is to solve the following optimization problem
\begin{prob}\label{p:Epoch_II_l}
\begin{equation}
\begin{array}{cll}
\mathop {\max} \limits_{\mathcal{S}_{1}} & {R}_{1} \left(\mathcal{S}_{1}, \mathcal{S}_{2} \right) \\
\text{s.t.} & \mathcal{S}_{1} \in \mathcal{F}_{1}\left(Q_1 - Q_{1}^{\text{III}}\right)
\end{array}
\end{equation}
\end{prob}
and the objective of seller 2 is to solve the following optimization problem
\begin{prob}\label{p:Epoch_II_s}
\begin{equation}
\begin{array}{cll}
\mathop {\max} \limits_{\mathcal{S}_{2}} & {R}_{2} \left(\mathcal{S}_{1}, \mathcal{S}_{2} \right) \\
\text{s.t.} & \mathcal{S}_{2} \in \mathcal{F}_{2}\left(Q_{2}^{\text{II}}\right).
\end{array}
\end{equation}
\end{prob}

For the non-cooperative game of the two sellers, a Nash equilibrium defines a strategy pair $(\mathcal{S}_{1}, \mathcal{S}_{2} )$ that a seller cannot earn more revenue by deviating from its strategy while keeping the other seller's strategy unchanged.
In other words, a Nash equilibrium should be a joint optimal solution of Problem \ref{p:Epoch_II_l} and Problem \ref{p:Epoch_II_s}.

Since the objective functions of Problem \ref{p:Epoch_II_l}
and Problem \ref{p:Epoch_II_s} are continuous and concave, and the feasible regions of the two sellers' strategies are convex, closed, bounded, and uncoupled\footnote{When the two feasible regions are independent from each other, we say that the two feasible regions are uncoupled.}, there exists at least one Nash equilibrium \cite{Rosen}.


\subsection{Spectrum Leasing Problem in Epoch III}
In Epoch III, only seller 1 is active in the spectrum market, and thus, monopoly spectrum leasing is performed. Once an amount of spectrum bandwidth is leased to secondary users, the spectrum can be used by secondary users until the end of Epoch III.

To maximize the revenue of seller $1$ in Epoch III, the following optimization problem should be solved.
\begin{prob}\label{p:Epoch_III}
\begin{equation} \label{e:V_function_Epoch_II}
\begin{array}{cll}
V\left(Q_{1}^{\text{III}}\right) \triangleq \mathop {\max} \limits_{\{d_{n, 1}|n \in \mathcal{N}_{\text{III}}\}} & \sum\limits_{n \in \mathcal{N}_{\text{III}}} { \left(C_0-C_1 d_{n, 1}\right)  d_{n,1} n} \\
\text{s.t.} & \sum \limits_{n \in \mathcal{N}_{\text{III}}} d_{n, 1} \leq Q_{1}^{\text{III}} \\
& d_{n,1} \geq 0, \forall n \in \mathcal{N}_{\text{III}}.
\end{array}
\end{equation}
\end{prob}
It can be seen that Problem \ref{p:Epoch_III} is a convex optimization problem, and thus, can be solved by existing numerical optimization methods.

From the perspective of seller $1$, it can adjust $Q_{1}^{\text{III}}$.
For a specific $Q_{1}^{\text{III}}$, the two sellers need to follow a Nash equilibrium in the non-cooperative game in Epoch II. Thus, the strategy of seller 1 can be written as $Q_{1}^{\text{III}}$ and $\mathcal{S}_{1}$, while the strategy of seller 2 can be written as $\mathcal{S}_{2}$.


When seller 1 reserves spectrum bandwidth $Q_{1}^{\text{III}}$ for Epoch III, it means that seller 1 would like to lease spectrum bandwidth ($Q_1-Q_{1}^{\text{III}}$) in Epoch II. Accordingly, we denote the revenue of seller 1 in Epoch II as $U(Q_1-Q_{1}^{\text{III}})$, a function of ($Q_1-Q_{1}^{\text{III}}$). Then for seller 1
to maximize its overall revenue,
the following optimization problem should be solved
\begin{prob}\label{p:high_level}
\begin{equation}
\begin{array}{cll}
\mathop {\max} \limits_{Q_{1}^{\text{III}}} & U\left(Q_1 - Q_{1}^{\text{III}}\right) + V\left(Q_{1}^{\text{III}}\right)  \\
\text{s.t.} & 0 \leq Q_{1}^{\text{III}} \leq  Q_1.\\
\end{array}
\end{equation}
\end{prob}

In the following,  in Section \ref{s:competetion} we find out Nash equilibria in Epoch II for a specific $Q_{1}^{\text{III}}$, and in Section \ref{s:monopoly} we select the value of $Q_{1}^{\text{III}}$ for seller 1.

\section{Nash Equilibria in the Non-Cooperative Game in Epoch II with Given $Q_{1}^{\text{III}}$}\label{s:competetion}

\subsection{Uniqueness of Nash Equilibrium in the Non-Cooperative Game in Epoch II when $|\mathcal{N}_{\text{III}}| \leq 12$}

We have the following theorem.
\begin{theorem} \label{lem:equilibrium_unique}
When $|\mathcal{N}_{\text{III}}| \leq 12$, there is only one Nash equilibrium for the non-cooperative game in Epoch II.
\end{theorem}

\begin{IEEEproof}

Define the vectorized strategy of seller 1 and seller 2 in Epoch II as
$\bm{x}_1=[d_{|\mathcal{N}_{\text{III}}|+ |\mathcal{N}_{\text{II}}|, 1},$ $d_{|\mathcal{N}_{\text{III}}|+ \left(|\mathcal{N}_{\text{II}}|-1\right), 1},$ $...,$ $d_{|\mathcal{N}_{\text{III}}|+ 1, 1}]^{T}$ and $\bm{x}_2=\left[d_{|\mathcal{N}_{\text{III}}|+ |\mathcal{N}_{\text{II}}|, 2}, d_{|\mathcal{N}_{\text{III}}|+ \left(|\mathcal{N}_{\text{II}}|-1\right), 2}, ..., d_{|\mathcal{N}_{\text{III}}|+ 1, 2}\right]^{T}$, respectively, in which $[\cdot]^T$ means transpose operation. The payoff function of seller 1 and seller 2 can be rewritten as $R_1\left(\mathcal{S}_1, \mathcal{S}_2\right)=R_1\left(\bm{x}_1, \bm{x}_2\right)$ and $R_2\left(\mathcal{S}_1, \mathcal{S}_2\right)=R_2\left(\bm{x}_1, \bm{x}_2\right)$, respectively.
Denote $\bm{x}=\left(\bm{x}_1^T, \bm{x}_2^T\right)^T$ and define
\begin{equation}
\sigma(\bm{x})= R_1\left(\bm{x}_1, \bm{x}_2\right) + R_2\left(\bm{x}_1, \bm{x}_2\right).
\end{equation}
Then the pseudo-gradient of $\sigma(\bm{x})$ can be given as
\begin{equation}
\bm{k}\left(\bm{x}\right)=\left[
\begin{array}{c}
\nabla_1 R_1\left(\bm{x}_1, \bm{x}_2\right) \\ \nabla_2 R_2\left(\bm{x}_1, \bm{x}_2\right)
\end{array}
\right]
\end{equation}
where $|\mathcal{N}_{\text{II}}|\times 1$ matrix $\nabla_1 R_1\left(\bm{x}_1, \bm{x}_2\right)$ is the gradient of $R_1\left(\bm{x}_1, \bm{x}_2\right)$ with respect to vector $\bm{x}_1$,
and $|\mathcal{N}_{\text{II}}|\times 1$ matrix $\nabla_2 R_2\left(\bm{x}_1, \bm{x}_2\right)$ is the gradient of
$R_2\left(\bm{x}_1, \bm{x}_2\right)$ with respect to vector $\bm{x}_2$.
According to Theorem 2 and Theorem 6 of \cite{Rosen}, the Nash equilibrium of the non-cooperative game in Epoch II is unique if the $2|\mathcal{N}_{\text{II}}|\times 2|\mathcal{N}_{\text{II}}|$ symmetric matrix $\bm{L}(\bm{x})=-\left[\bm{K}(\bm{x})+\bm{K}^T(\bm{x})\right]$ is positive definite, where $\bm{K}(\bm{x})$ is
the Jacobian of $\bm{k}(\bm{x})$ with respect to $\bm{x}$.
After some math manipulation, the matrix $\bm{L}(\bm{x})$ can be written as the following form
\begin{equation}
\bm{L}(\bm{x}) = \left[
\begin{array}{c}
\bm{L}_{11}(\bm{x}) \\ \bm{L}_{21}(\bm{x})
\end{array} \
\begin{array}{c}
\bm{L}_{12}(\bm{x}) \\ \bm{L}_{22}(\bm{x})
\end{array}
\right]
\end{equation}
where
$\bm{L}_{11}(\bm{x})=\text{Diag}\big(4C_1(|\mathcal{N}_{\text{III}}|+|\mathcal{N}_{\text{II}}|), 4C_1(|\mathcal{N}_{\text{III}}|+|\mathcal{N}_{\text{II}}|-1), ..., 4C_1(|\mathcal{N}_{\text{III}}|+1)\big)$,
$\bm{L}_{12}(\bm{x})=\bm{L}_{21}(\bm{x})=\text{Diag}\big(C_1(|\mathcal{N}_{\text{III}}|+2|\mathcal{N}_{\text{II}}|), C_1(|\mathcal{N}_{\text{III}}|+2|\mathcal{N}_{\text{II}}|-2), ..., C_1(|\mathcal{N}_{\text{III}}|+2)\big)$,
and
$\bm{L}_{22}(\bm{x})=\text{Diag}\big(4C_1|\mathcal{N}_{\text{II}}|, 4C_1(|\mathcal{N}_{\text{II}}|-1), ..., 4C_1\big)$. Here $\text{Diag}(\cdot \cdot \cdot)$ means a diagonal matrix with all diagonal elements listed in $(\cdot \cdot \cdot)$. The matrix $\bm{L}(\bm{x})$ can be guaranteed to be positive definite, if the leading principal minors are all positive \cite{Meryer}, i.e., the determinant of $m\times m$ upper-left submatrix of $\bm{L}(\bm{x})$ is larger than 0 for $m=1, 2, ..., 2|\mathcal{N}_{\text{II}}|$.
Since there is
\begin{equation*}
\text{Det} \left(
\left[
\begin{array}{c}
\bm{A} \\ \bm{C}
\end{array} \
\begin{array}{c}
\bm{B} \\ \bm{D}
\end{array}
\right] \right)
= \text{Det}\left(\bm{A}\right)\text{Det}\left(\bm{D} - \bm{C} \bm{A}^{-1} \bm{B}\right)
\end{equation*}
when matrix $\bm{A}$ is invertible \cite{Zhangxianda},
the determinant of $m\times m$ upper-left submatrix of $\bm{L}(\bm{x})$ is larger than 0 for $m=1, 2, ..., 2|\mathcal{N}_{\text{II}}|$ when the following inequalities hold
\begin{equation}
12\left(|\mathcal{N}_{\text{II}}|-k\right)^2
+ 12 |\mathcal{N}_{\text{III}}|\left(|\mathcal{N}_{\text{II}}| - k\right) - |\mathcal{N}_{\text{III}}|^2 > 0, \forall k=0, 1, ..., \left(|\mathcal{N}_{\text{II}}|-1\right),
\end{equation}
i.e., when
\begin{equation} \label{e:principal_minor_larger_than_zero}
\frac{\left(|\mathcal{N}_{\text{II}}|-k\right)}{|\mathcal{N}_{\text{III}}|} > \left(-\frac{1}{2}+ \frac{1}{\sqrt{3}}\right), \forall k \in 0, 1, ..., \left(|\mathcal{N}_{\text{II}}|-1\right).
\end{equation}
The inequalities in (\ref{e:principal_minor_larger_than_zero}) hold if
\begin{equation}
|\mathcal{N}_{\text{III}}|< \frac{1}{-\frac{1}{2}+ \frac{1}{\sqrt{3}}}=12.9282.
\end{equation}

This completes the proof.
\end{IEEEproof}

As the number of stages in Epoch III is normally limited, it is very likely that the value of $|\mathcal{N}_{\text{III}}|$ is bounded by 12, and thus, Nash equilibrium of the non-cooperative game in Epoch II is unique. Nevertheless, in next subsection, we show how to find Nash equilibria in the non-cooperative game in Epoch II without constraint $|\mathcal{N}_{\text{III}}|\le 12$ (i.e., Nash equilibrium may or may not be unique).

\subsection{Finding Nash Equilibria in the Non-Cooperative Game in Epoch II}


As aforementioned, a Nash equilibrium of the non-cooperative game in Epoch II is a joint optimal solution of Problem \ref{p:Epoch_II_l} and Problem \ref{p:Epoch_II_s}.
As both Problem \ref{p:Epoch_II_l} and Problem \ref{p:Epoch_II_s} are convex problems and satisfy the Slater's condition, KKT condition is a sufficient and necessary condition for optimal solution for each problem \cite{Boyd,Xing1, Xing2}.

For the ease of presentation, we denote
$Q_{1}^{\text{II}_c} = Q_1 - Q_{1}^{\text{III}}$ as the spectrum bandwidth amount that seller 1 would like to lease to secondary users in Epoch II.
For Problem \ref{p:Epoch_II_l}, the KKT condition is
\begin{subequations} \label{e:KKT_II_l}
\begin{eqnarray}
2 C_1 n d_{n,1} \! -\! \left(C_0 \!-\! C_1 d_{n,2}\right) n + \lambda -\mu_n =0, ~\forall n\in \mathcal{N}_{\text{II}} \label{e:KKT_l_utility} \\
\lambda\left(\sum \limits_{n \in \mathcal{N}_{\text{II}}} d_{n, 1} - Q_{1}^{\text{II}_c} \right)=0 \\
\mu_n d_{n,1}=0, ~\forall n\in \mathcal{N}_{\text{II}} \label{e:KKT_l_d_mu_zero} \\
\sum \limits_{n \in \mathcal{N}_{\text{II}}} d_{n, 1} \leq Q_{1}^{\text{II}_c} \label{e:KKT_II_l_demand_sum} \\
d_{n,1} \geq 0, \forall n \in \mathcal{N}_{\text{II}} \\
\lambda \geq 0; \mu_n \geq 0, ~\forall n\in \mathcal{N}_{\text{II}}
\end{eqnarray}
\end{subequations}
where $\lambda$ and $\mu_n$ are Lagrange multipliers associated with the constraints $\sum \limits_{n \in \mathcal{N}_{\text{II}}} d_{n, 1} \leq Q_{1}^{\text{II}_c}$ and $d_{n,1} \geq 0$, respectively.

For Problem \ref{p:Epoch_II_s}, the KKT condition is
\begin{subequations} \label{e:KKT_II_s}
\begin{eqnarray}
2 C_1 \left(n - |\mathcal{N}_{\text{III}}|\right)d_{n,2}  - \left(C_0 - C_1 d_{n,1}\right) \left(n - |\mathcal{N}_{\text{III}}|\right) \notag \\+ \zeta -\nu_n =0, ~\forall n\in \mathcal{N}_{\text{II}} \label{e:KKT_s_utility} \\
\zeta \left(\sum \limits_{n \in \mathcal{N}_{\text{II}}} d_{n, 2} - Q_{2}^{\text{II}} \right)=0 \\
\nu_n d_{n,2}=0, ~\forall n\in \mathcal{N}_{\text{II}}  \label{e:KKT_s_d_nu_zero}\\
\sum \limits_{n \in \mathcal{N}_{\text{II}}} d_{n, 2} \leq Q_{2}^{\text{II}} \label{e:KKT_II_s_demand_sum}\\
d_{n,2} \geq 0, \forall n \in \mathcal{N}_{\text{II}} \\
\zeta \geq 0; \nu_n \geq 0, ~\forall n\in \mathcal{N}_{\text{II}}
\end{eqnarray}
\end{subequations}
where $\zeta$ and $\nu_n$ are Lagrange multipliers associated with the constraints $\sum \limits_{n \in \mathcal{N}_{\text{II}}} d_{n, 2} \leq Q_{2}^{\text{II}}$ and $d_{n,2} \geq 0$, respectively.

To get Nash equilibrium of the non-cooperative game in Epoch II, the equations (\ref{e:KKT_II_l}) and (\ref{e:KKT_II_s}) should be solved jointly.

We have two properties for the joint optimal solution:
\begin{itemize}
\item {\bf Property 1:} Equality should hold in (\ref{e:KKT_II_l_demand_sum}) and (\ref{e:KKT_II_s_demand_sum}) (in other words, we have $\sum \limits_{n \in \mathcal{N}_{\text{II}}} d_{n, 1} = Q_{1}^{\text{II}_c}$ and $\sum \limits_{n \in \mathcal{N}_{\text{II}}} d_{n, 2} = Q_{2}^{\text{II}}$).

\item {\bf Property 2:} If $d_{n,1}>0$ ($n \in \mathcal{N}_{\text{II}}$), then we have $\mu_n=0$; if $d_{n,2}>0$, then we have $\nu_{n}=0$.
\end{itemize}
Property 1 is due to the facts that the objective function of Problem \ref{p:Epoch_II_l} is a monotonically increasing function of $d_{n,1}$ ($n\in \mathcal{N}_{\text{II}}$) and that the objective function of Problem \ref{p:Epoch_II_s} is a monotonically increasing function of $d_{n,2}$ ($n\in \mathcal{N}_{\text{II}}$). Property 2 can be obtained directly from the equalities (\ref{e:KKT_l_d_mu_zero}) and (\ref{e:KKT_s_d_nu_zero}).

Next, we try to find the expressions of $d_{n,1}$ and $d_{n,2}$ by solving (\ref{e:KKT_II_l}) and (\ref{e:KKT_II_s}).

From the equalities (\ref{e:KKT_l_utility}) and (\ref{e:KKT_s_utility}), $d_{n,1}$ and $d_{n,2}$ for $n \in \mathcal{N}_{\text{II}}$ can be expressed as
\begin{equation} \label{e:dn1_by_dn2_II}
d_{n,1}= \frac{\left(C_0 -C_1 d_{n,2}\right) n -\lambda + \mu_n }{2 C_1 n },
\end{equation}
\begin{equation} \label{e:dn2_by_dn1_II}
d_{n,2} = \frac{\left(C_0 -C_1 d_{n,1}\right) \left(n - |\mathcal{N}_{\text{III}}|\right) - \zeta + \nu_n }{2C_1  \left(n - |\mathcal{N}_{\text{III}}|\right)},
\end{equation}
from which we have
\begin{equation} \label{e:dn1_general_II}
d_{n,1} = \frac{2\left(C_0 n - \lambda + \mu_n \right)}{3C_1n} -\frac{C_0 \left(n - |\mathcal{N}_{\text{III}}|\right) -\zeta + \nu_n }{3 C_1 \left(n - |\mathcal{N}_{\text{III}}|\right)},
\end{equation}
\begin{equation} \label{e:dn2_general_II}
d_{n,2} = -\frac{C_0 n - \lambda + \mu_n }{3C_1n} + \frac{2 \left( C_0 \left(n - |\mathcal{N}_{\text{III}}|\right) -\zeta + \nu_n \right) }{3 C_1 \left(n - |\mathcal{N}_{\text{III}}|\right)}.
\end{equation}

Define $\mathcal{Z}_{1}=\{n| d_{n,1}>0, d_{n,2}>0, n \in \mathcal{N}_{\text{II}}\}$, $\mathcal{Z}_{2}=\{n| d_{n,1}>0, d_{n,2}=0, n\in \mathcal{N}_{\text{II}}\}$, $\mathcal{Z}_{3}=\{n| d_{n,1}=0, d_{n,2}>0, n\in \mathcal{N}_{\text{II}}\}$ and $\mathcal{Z}_{4}=\{n| d_{n,1}=0, d_{n,2}=0, n\in \mathcal{N}_{\text{II}}\}$. Then \{$\mathcal{Z}_1$, $\mathcal{Z}_2$, $\mathcal{Z}_3$, $\mathcal{Z}_4$\} constitutes a decomposition of the set $\mathcal{N}_{\text{II}}$, which means that $\mathcal{Z}_1\bigcup\mathcal{Z}_2\bigcup \mathcal{Z}_3\bigcup\mathcal{Z}_4=\mathcal{N}_{\text{II}}$ and $\mathcal{Z}_i\bigcap\mathcal{Z}_j= \emptyset$ for $i \neq j$ and $i,j \in \{1,2,3,4\}$. Totally there are $2^{2|\mathcal{N}_\text{II}|}$ decompositions.

Next we find out the expressions of $d_{n,1}$ and $d_{n,2}$ for a specific decomposition \{$\mathcal{Z}_1$, $\mathcal{Z}_2$, $\mathcal{Z}_3$, $\mathcal{Z}_4$\}.

From Property 1, we have
\[\sum_{n\in \mathcal{Z}_1} d_{n,1} + \sum_{n\in \mathcal{Z}_2} d_{n,1} = Q_{1}^{\text{II}_c},\]
\[\sum_{n\in \mathcal{Z}_1} d_{n,2} + \sum_{n\in \mathcal{Z}_3} d_{n,2} = Q_{2}^{\text{II}}.\]
In the two equations, substituting the expressions of $d_{n,1}$ and $d_{n,2}$ in (\ref{e:dn1_general_II}) and (\ref{e:dn2_general_II}) for $n \in \mathcal{Z}_1$, substituting the expressions of $d_{n,1}$ and $d_{n,2}$ in (\ref{e:dn1_by_dn2_II}) and (\ref{e:dn2_by_dn1_II}) for $n \in \mathcal{Z}_2$ and $n\in \mathcal{Z}_3$, and using Property 2, we have the following equations:
\begin{equation} \label{e:lambda_zeta_II_general}
\begin{array}{ll}
& -A_{11} \lambda + A_{12} \zeta = Q_1^{\text{II}_c} - \sum \limits_{n \in \mathcal{Z}_1} \frac{C_0}{3C_1} - \sum \limits_{n \in \mathcal{Z}_2} \frac{C_0}{2C_1} \\
& A_{21} \lambda - A_{22} \zeta = Q_2^{\text{II}} - \sum \limits_{n \in \mathcal{Z}_1} \frac{C_0}{3C_1} - \sum \limits_{n \in \mathcal{Z}_3} \frac{C_0}{2C_1} \\
\end{array}
\end{equation}
where
\begin{equation} \label{e:A_11}
A_{11} = \sum \limits_{n\in \mathcal{Z}_1} \frac{2}{3C_1 n} + \sum \limits_{n \in \mathcal{Z}_2} \frac{1}{2C_1 n},
\end{equation}
\begin{equation} \label{e:A_12}
A_{12} = \sum \limits_{n \in \mathcal{Z}_1} \frac{1}{3 C_1  \left(n - |\mathcal{N}_{\text{III}}|\right)},
\end{equation}
\begin{equation} \label{e:A_21}
A_{21} = \sum  \limits_{n \in \mathcal{Z}_1} \frac{1}{3 C_1  n},
\end{equation}
\begin{equation} \label{e:A_22}
A_{22} = \sum \limits_{n \in \mathcal{Z}_1} \frac{2}{3C_1  \left(n - |\mathcal{N}_{\text{III}}|\right)} + \sum \limits_{n \in \mathcal{Z}_3} \frac{1}{2C_1 \left(n - |\mathcal{N}_{\text{III}}|\right)}.
\end{equation}
Note that $A_{11}$, $A_{12}$, $A_{21}$ and $A_{22}$ are all larger than zero.
According to the equations in (\ref{e:lambda_zeta_II_general}), the Lagrange multipliers $\lambda$ and $\zeta$ can be expressed as
\begin{equation} \label{e:lambda_final}
\begin{array}{lr}
\lambda  =& -\frac{A_{22}}{A_{11}A_{22} - A_{21}A_{12}} \left(Q_1^{\text{II}_c} - \sum \limits_{n \in \mathcal{Z}_1} \frac{C_0}{3C_1} - \sum \limits_{n \in \mathcal{Z}_2} \frac{C_0}{2C_1}\right) \\
& - \frac{A_{12}}{A_{11}A_{22} - A_{21}A_{12}} \left( Q_2^{\text{II}} - \sum \limits_{n \in \mathcal{Z}_1} \frac{C_0}{3C_1} - \sum \limits_{n \in \mathcal{Z}_3} \frac{C_0}{2C_1}\right),
\end{array}
\end{equation}
\begin{equation} \label{e:zeta_final}
\begin{array}{lr}
\zeta =& -\frac{A_{21}}{A_{11}A_{22} - A_{21}A_{12}} \left(Q_1^{\text{II}_c} - \sum \limits_{n \in \mathcal{Z}_1} \frac{C_0}{3C_1} - \sum \limits_{n \in \mathcal{Z}_2} \frac{C_0}{2C_1}\right) \\
& - \frac{A_{11}}{A_{11}A_{22} - A_{21}A_{12}} \left( Q_2^{\text{II}} - \sum \limits_{n \in \mathcal{Z}_1} \frac{C_0}{3C_1} - \sum \limits_{n \in \mathcal{Z}_3} \frac{C_0}{2C_1}\right).
\end{array}
\end{equation}

With the aid of Property 2 and using equations (\ref{e:dn1_by_dn2_II}), (\ref{e:dn2_by_dn1_II}), (\ref{e:dn1_general_II}), and (\ref{e:dn2_general_II}),
the closed-form expressions of $d_{n,1}$ and $d_{n,2}$ for $n \in \mathcal{N}_{\text{II}}$ are given as follows:
\begin{equation} \label{e:dn1_II_final}
d_{n,1}=
\begin{cases}
\frac{2\left(C_0 n - \lambda  \right)}{3C_1 n} -\frac{C_0 \left(n - |\mathcal{N}_{\text{III}}|\right)  -\zeta }{3 C_1 \left(n - |\mathcal{N}_{\text{III}}|\right)} &\text{if } n \in \mathcal{Z}_1 \\
\frac{C_0 n -\lambda }{2 C_1 n} &\text{if } n \in \mathcal{Z}_2 \\
0 &\text{if } n \in \mathcal{Z}_3 \bigcup \mathcal{Z}_4\\
\end{cases}
\end{equation}
\begin{equation} \label{e:dn2_II_final}
d_{n,2}=
\begin{cases}
-\frac{C_0 n - \lambda }{3C_1 n} + \frac{2 \left( C_0 \left(n - |\mathcal{N}_{\text{III}}|\right) -\zeta \right) }{3 C_1 \left(n - |\mathcal{N}_{\text{III}}|\right)} &\text{if } n \in \mathcal{Z}_{1} \\
\frac{C_0 \left(n - |\mathcal{N}_{\text{III}}|\right) - \zeta }{2C_1  \left(n - |\mathcal{N}_{\text{III}}|\right)}  &\text{if } n \in \mathcal{Z}_3 \\
0 &\text{if } n \in \mathcal{Z}_2 \bigcup \mathcal{Z}_4\\
\end{cases}
\end{equation}
where $\lambda$ and $\zeta$ are given in (\ref{e:lambda_final}) and (\ref{e:zeta_final}), respectively.

By now, given the decomposition $\{\mathcal{Z}_1, \mathcal{Z}_2, \mathcal{Z}_3, \mathcal{Z}_4\}$, expressions of $d_{n,1}$ and $d_{n,2}$ for $n \in \mathcal{N}_{\text{II}}$ are derived. To guarantee that every equality or inequality in (\ref{e:KKT_II_l}) and (\ref{e:KKT_II_s}) is satisfied, a feasibility check is further required, which is given as follows:
\begin{enumerate}
\item $\lambda$ and $\zeta$, which can be calculated from (\ref{e:lambda_final}) and (\ref{e:zeta_final}), are non-negative.
\item $d_{n,1}$ and $d_{n,2}$, which are calculated from (\ref{e:dn1_II_final}) and (\ref{e:dn2_II_final}), are non-negative for $n\in \mathcal{N}_{\text{II}}$.
\item $\mu_n$ and $\nu_n$, which can be calculated from (\ref{e:dn1_by_dn2_II}) and (\ref{e:dn2_by_dn1_II}) given the obtained $d_{n,1}$, $d_{n,2}$, $\lambda$ and $\zeta$, are non-negative for $n \in \mathcal{N}_{\text{II}}$.
\end{enumerate}
If the above feasibility check passes, the decomposition $\{\mathcal{Z}_1, \mathcal{Z}_2, \mathcal{Z}_3, \mathcal{Z}_4\}$ is said to be {\it feasible}, and the derived $d_{n,1}$ and $d_{n,2}$ expressions in (\ref{e:dn1_II_final}) and (\ref{e:dn2_II_final}) for $n \in \mathcal{N}_{\text{II}}$ given the decomposition $\{\mathcal{Z}_1, \mathcal{Z}_2, \mathcal{Z}_3, \mathcal{Z}_4\}$ is a Nash equilibrium of the non-cooperative game in Epoch II.

For the set $\mathcal{N}_{\text{II}}$, there are $2^{2|\mathcal{N}_\text{II}|}$ possible decompositions.
To find all Nash equilibria of the game in Epoch II, an exhaustive search of all $2^{2|\mathcal{N}_\text{II}|}$ decompositions is required. As the number of stages in Epoch II is normally very limited, and the calculations in checking feasibility of each decomposition are simple, an exhaustive search of all $2^{2|\mathcal{N}_\text{II}|}$ decompositions is considered to be acceptable. In addition, the following theorem is helpful in reducing the complexity in the exhaustive search.

\begin{theorem}\label{lem:complexity_reduction}
For a feasible decomposition, if there exists a stage (say stage $n$) in $\mathcal{Z}_4$ (i.e., $d_{n,1}=d_{n,2}=0$), then all stages with a lower index in Epoch II should belong to $\mathcal{Z}_4$.
\end{theorem}
\begin{IEEEproof}
We use the proof by contradiction. In the Nash equilibrium of the decomposition, suppose there is $n\dag$ satisfying $n\dag < n, n\dag \in \mathcal{N}_{\text{II}} \backslash \mathcal{Z}_4$. We first assume that $n\dag \in \mathcal{Z}_2$, which indicates that $d_{n\dag,1}>0, d_{n\dag,2}=0$. Then the total revenue collected over stage $n$ and stage $n\dag$ by seller 1 is $\left(C_0 -C_1 d_{n\dag,1}\right)d_{n\dag,1}n{\dag}$. By interchanging seller 1's offered spectrum bandwidth amounts in stage $n$ and stage $n\dag$, the total revenue that seller 1 collects in stages $n$ and $n\dag$ becomes $\left(C_0 -C_1 d_{n\dag,1}\right)d_{n\dag,1}n$, which is larger than $\left(C_0 -C_1 d_{n\dag,1}\right)d_{n\dag,1}n\dag$ since $n\dag < n$. This contradicts the definition of Nash equilibrium.

Similarly, $n\dag \in \mathcal{Z}_1$ or $n\dag \in \mathcal{Z}_3$ also leads to a contradiction.

This completes the proof.
\end{IEEEproof}

Remark: Theorem \ref{lem:complexity_reduction} shows that in a feasible decomposition, if $\mathcal{Z}_4$ is not empty, then it contains consecutive stages until the end of Epoch II. Therefore, in the exhaustive search of all possible decompositions, we can skip those decompositions in which $\mathcal{Z}_4$ contains non-consecutive stages or does not last until the end of Epoch II. Thus, the number of decompositions that we should check reduces from $2^{|2\mathcal{N}_{\text{II}}|}$ to $\sum_{i=0}^{|\mathcal{N}_{\text{II}}|} 3^i$.

So far all Nash equilibria of the non-cooperative game in Epoch II have been found. If there exists only one unique Nash equilibrium (e.g., when $|\mathcal{N}_{\text{III}}| \leq 12$), then both sellers follow the unique Nash equilibrium. If there are two or more Nash equilibria, the two sellers need to select one Nash equilibrium to follow. Here it is assumed that the two sellers agree to follow the Nash equilibrium that maximizes the minimum unit-bandwidth revenue of the two sellers. Here for seller 1, its unit-bandwidth revenue is the ratio of its total revenue in Epoch II to $Q_1^{\text{II}_c}$; for seller 2, its unit-bandwidth revenue is the ratio of its total revenue in Epoch II to $Q_2^{\text{II}}$.

\section{Total Revenue Maximization for seller 1} \label{s:monopoly}
In the previous section, we have found the strategies of the two sellers in Epoch II with a specific $Q_{1}^{\text{III}}$ (the bandwidth that seller 1 reserves for Epoch III). Now, we try to solve Problem \ref{p:high_level}, i.e., find out the optimal value of $Q_{1}^{\text{III}}$ that maximizes seller 1's total revenue. A method could be: 1) for each possible value of $Q_{1}^{\text{III}}$, search all possible Nash equilibria, find the Nash equilibrium that maximizes the minimum unit-bandwidth revenue of the two sellers, and calculate the revenue that seller 1 can earn during its leasing period; 2) compare the revenue values that seller 1 can earn during its leasing period with different $Q_{1}^{\text{III}}$, and select the optimal $Q_{1}^{\text{III}}$ that makes seller 1 earn the most revenue. However, the complexity of the method is huge, due to the infinite number of values of  $Q_{1}^{\text{III}}\in [0, Q_1]$. Thus, we target at an approximation method to select $Q_{1}^{\text{III}}$. When $Q_{1}^{\text{III}}=x$, $U(Q_1-x)$ and $V(x)$ given in (\ref{e:V_function_Epoch_II}) are the revenue of seller 1 in Epoch II and Epoch III, respectively. To select $x$ (i.e., $Q_{1}^{\text{III}}$), we need to evaluate how $V\left(x\right)$ and $U\left(Q_1-x\right)$ change when $x$ varies.

\begin{lem} \label{lem:V_increasing_concave}
The function $V(x)$ is an increasing and concave function with $x$.
\end{lem}
\begin{IEEEproof}
The proof follows a similar procedure to the proof of Lemma 6 of \cite{Rongfei10}.
%
\end{IEEEproof}

Now we evaluate function $U(Q_1-x)$ when $x$ varies. To evaluate $U(Q_1-x)$ for a specific decomposition $\left\{\mathcal{Z}_1, \mathcal{Z}_2, \mathcal{Z}_3, \mathcal{Z}_4\right\}$, we need to know $d_{n,1}$ and $d_{n,2}$ ($n\in \mathcal{N}_\text{II}$) in the Nash equilibrium corresponding to the decomposition. Therefore, next we show how $d_{n,1}$ and $d_{n,2}$ change when $x$ varies.

Consider a decomposition $\left\{\mathcal{Z}_1, \mathcal{Z}_2, \mathcal{Z}_3, \mathcal{Z}_4\right\}$. Consider two $Q_{1}^{\text{II}_c}$ values (recalling that $Q_{1}^{\text{II}_c} = Q_1 - Q_{1}^{\text{III}}$): $Q^{\dag}$ and $Q^{\ddag}$, with $Q^{\dag} \leq Q^{\ddag}$. Assume the decomposition $\left\{\mathcal{Z}_1, \mathcal{Z}_2, \mathcal{Z}_3, \mathcal{Z}_4\right\}$ is feasible for both $Q_{1}^{\text{II}_c} $ values. For the decomposition, denote the corresponding Nash equilibrium when $Q_{1}^{\text{II}_c} =Q^{\dag}$ as
\begin{equation*}
\left(\mathcal{S}_1^{\dag}, \mathcal{S}_2^{\dag}\right) \triangleq \left(\left\{d_{n,1}^{\dag}|n \in \mathcal{N}_{\text{II}}\right\}, \left\{d_{n,2}^{\dag}|n \in \mathcal{N}_{\text{II}}\right\}\right),
\end{equation*}
and the corresponding Nash equilibrium when $Q_{1}^{\text{II}_c} =Q^{\ddag}$ as
\begin{equation*}
\left(\mathcal{S}_1^{\ddag}, \mathcal{S}_2^{\ddag}\right) \triangleq \left(\left\{d_{n,1}^{\ddag}|n \in \mathcal{N}_{\text{II}}\right\}, \left\{d_{n,2}^{\ddag}|n \in \mathcal{N}_{\text{II}}\right\}\right).
\end{equation*}
Then the following lemmas can be expected.
\begin{lem} \label{lem:d_1_mono}
For seller 1, $d_{n,1}^{\dag} \leq d_{n,1}^{\ddag}$ for $n \in \mathcal{Z}_2$, and $d_{n,1}^{\dag} = d_{n,1}^{\ddag}=0$ for $n  \in \mathcal{Z}_3 \bigcup \mathcal{Z}_4$.
\end{lem}
\begin{IEEEproof}
By the definitions of set $\mathcal{Z}_3$ and $\mathcal{Z}_4$, seller 1 does not offer spectrum bandwidth to be leased in stages in $\mathcal{Z}_3$ and $\mathcal{Z}_4$, and thus, $d_{n,1}^{\dag} = d_{n,1}^{\ddag}=0$ for $n  \in \mathcal{Z}_3 \bigcup \mathcal{Z}_4$.

For $n \in \mathcal{Z}_2$, with the aid of (\ref{e:dn1_II_final}) and (\ref{e:lambda_final}), we have
\begin{equation}
\begin{array}{lll}
d_{n, 1}^{\dag} - d_{n, 1}^{\ddag} &= \frac{C_0 A_{22}}{2C_1 n \left(A_{11}A_{22} - A_{21}A_{12}\right)} \left(Q^{\dag} -Q^{\ddag}\right) \\
& \leq 0
\end{array}
\end{equation}
in which the inequality comes from $A_{22} \geq 0$, $Q^{\dag} \leq Q^{\ddag}$, and $\left(A_{11}A_{22} - A_{21}A_{12}\right) >0$ according to (\ref{e:A_11}), (\ref{e:A_12}), (\ref{e:A_21}), and (\ref{e:A_22}).


This completes the proof.
\end{IEEEproof}

\begin{lem} \label{lem:d_2_mono}
For seller 2, $d_{n,2}^{\dag} \leq d_{n,2}^{\ddag}$ for $n \in \mathcal{Z}_3$, and $d_{n,2}^{\dag} = d_{n,2}^{\ddag}=0$ for $n  \in \mathcal{Z}_2 \bigcup \mathcal{Z}_4$.
\end{lem}
\begin{IEEEproof}
The proof is similar to the proof for Lemma \ref{lem:d_1_mono}, and thus, is omitted here.
\end{IEEEproof}

\begin{theorem} \label{lem:U_x_differential_mono}
If a decomposition $\left\{\mathcal{Z}_1, \mathcal{Z}_2, \mathcal{Z}_3, \mathcal{Z}_4\right\}$ is feasible when $Q_{1}^{\text{III}}=x \in \mathcal{I}$ where $\mathcal{I} \subseteq [0,Q_1] $ is an interval,
then when the Nash equilibrium corresponding to the decomposition is followed by the two sellers in Epoch II, seller 1's revenue $U(Q_1-x)$ in Epoch II can be written as $U(Q_1-x)=G(x)-H(x)$ where $G(x)$ and $H(x)$ are monotonically increasing functions with respect to $x\in \mathcal{I}$.
\end{theorem}
\begin{IEEEproof}
Suppose the Nash equilibrium corresponding to the decomposition $\left\{\mathcal{Z}_1, \mathcal{Z}_2, \mathcal{Z}_3, \mathcal{Z}_4\right\}$ is
$(\left\{d_{n,1}|n \in \mathcal{N}_{\text{II}}\right\}$, $\left\{d_{n,2}|n \in \mathcal{N}_{\text{II}}\right\})$.
Then $U(Q_1-x)$ can be written as
\begin{equation} \label{e:U_differential_proof}
\begin{array}{l}
U(Q_1-x) \\=\sum \limits_{n\in \mathcal{N}_\text{II}} \left(C_0-C_1 \left(d_{n,1} + d_{n,2} \right)\right) d_{n,1} n \\
\overset{(a)}=  \sum \limits_{n \in \mathcal{Z}_1} \left(C_0-C_1 \left(d_{n,1} + d_{n,2} \right)\right) d_{n,1} n \\
~~~~~~~~ + \sum \limits_{n \in \mathcal{Z}_2} \left(C_0-C_1 d_{n,1} \right) d_{n,1} n \\
\overset{(b)}=  \sum \limits_{n \in \mathcal{Z}_1} \left(\frac{C_0}{3} + \frac{\zeta }{3 \left(n-|\mathcal{N}_{\text{III}}|\right)} + \frac{\lambda }{3 n}\right) \Big(\frac{\zeta }{3 C_1 \left(n-|\mathcal{N}_{\text{III}}|\right)}-\frac{2 \lambda }{3 C_1 n}\\~~~~~~~~+\frac{C_0}{3 C_1}\Big) n
+ \sum \limits_{n \in \mathcal{Z}_2} \left(C_0-C_1 d_{n,1} \right) d_{n,1} n \\
= \sum \limits_{n \in \mathcal{Z}_1} \left(\frac{\zeta ^2}{9 C_1 \left(n-|\mathcal{N}_{\text{III}}|\right){}^2} +\frac{2 C_0 \zeta }{9 C_1 \left(n-|\mathcal{N}_{\text{III}}|\right)}+\frac{C_0^2}{9 C_1}\right) n
\\
 ~~~~~~~~- {\sum \limits_{n \in \mathcal{Z}_1}} \left(\frac{\zeta  \lambda}{9 C_1 n \left(n-|\mathcal{N}_{\text{III}}|\right)} + \frac{2 \lambda ^2}{9 C_1 n^2} + \frac{C_0 \lambda }{9 C_1 n}\right) n \\~~~~~~~~+ \sum \limits_{n \in \mathcal{Z}_2} \left(C_0-C_1 d_{n,1} \right) d_{n,1} n
\end{array}
\end{equation}
where $(a)$ holds since $d_{n,1}=0$ for $n \in \mathcal{Z}_3 \bigcup \mathcal{Z}_4$ and $d_{n,2}=0$ for $n \in \mathcal{Z}_2$, and $(b)$ can be  obtained by substituting $d_{n,1}$ and $d_{n,2}$ according to (\ref{e:dn1_II_final}) and (\ref{e:dn2_II_final}).

As the decomposition $\left\{\mathcal{Z}_1, \mathcal{Z}_2, \mathcal{Z}_3, \mathcal{Z}_4\right\}$ is feasible, $\lambda$ and $\zeta$ are non-negative. Additionally, from (\ref{e:lambda_final}) and (\ref{e:zeta_final}), it can be seen that $\lambda$ and $\zeta$ are monotonically decreasing with $Q_1^{\text{II}_c}$, i.e., $(Q_1-x)$. So in the expression (\ref{e:U_differential_proof}), both the term
$\sum \limits_{n \in \mathcal{Z}_1} \left(\frac{\zeta ^2}{9 C_1 \left(n-|\mathcal{N}_{\text{III}}|\right){}^2} +\frac{2 C_0 \zeta }{9 C_1 \left(n-|\mathcal{N}_{\text{III}}|\right)}+\frac{C_0^2}{9 C_1}\right) n$
and the term
${\sum \limits_{n \in \mathcal{Z}_1}} \left(\frac{\zeta  \lambda}{9 C_1 n \left(n-|\mathcal{N}_{\text{III}}|\right)} + \frac{2 \lambda ^2}{9 C_1 n^2} + \frac{C_0 \lambda }{9 C_1 n}\right) n$
are monotonically decreasing with $Q_1^{\text{II}_c}$, and thus, are monotonically increasing with $x$ (as $Q_1^{\text{II}_c}=Q_1-x$).
It can be also checked that the term
$\sum \limits_{n \in \mathcal{Z}_2} \left(C_0-C_1 d_{n,1} \right) d_{n,1} n$
in (\ref{e:U_differential_proof}) is a monotonically increasing function with respect to $Q_1^{\text{II}_c}$ (since the function $\left(C_0 -C_1y\right)y$ is monotonically increasing with $y$ and $d_{n,1}$ grows with $Q_1^{\text{II}_c}$ [from Lemma \ref{lem:d_1_mono}]), and thus, is a monotonically decreasing function with respect to $x$.

Define
\begin{equation}
G(x)= \! \sum \limits_{n \in \mathcal{Z}_1} \!\left(\!\frac{\zeta ^2}{9 C_1 \left(n\!-\!|\mathcal{N}_{\text{III}}|\right){}^2} \!+\!\frac{2 C_0 \zeta }{9 C_1 \left(n\!-\!|\mathcal{N}_{\text{III}}|\right)}+\frac{C_0^2}{9 C_1}\right)\! n
\end{equation}
and
\begin{equation}
H(x) =  {\sum \limits_{n \in \mathcal{Z}_1}} \left(\frac{\zeta  \lambda}{9 C_1 n \left(n-|\mathcal{N}_{\text{III}}|\right)} + \frac{2 \lambda ^2}{9 C_1 n^2} + \frac{C_0 \lambda }{9 C_1 n}\right) n  - \sum \limits_{n \in \mathcal{Z}_2} \left(C_0-C_1 d_{n,1} \right) d_{n,1} n.
\end{equation}
It can be seen that $U(Q_1-x)=G(x)-H(x)$, and both $G(x)$ and $H(x)$ monotonically increase with $x$.

This completes the proof.
\end{IEEEproof}

In Lemma \ref{lem:d_1_mono}, Lemma \ref{lem:d_2_mono}, and Theorem \ref{lem:U_x_differential_mono},
it is assumed that the decomposition $\left\{\mathcal{Z}_1, \mathcal{Z}_2, \mathcal{Z}_3, \mathcal{Z}_4\right\}$ is feasible for $x=Q_1-Q^{\dag}$, $x=Q_1-Q^{\ddag}$ or $x \in \mathcal{I}$. The next theorem will answer the following question: If a decomposition is feasible for a specific value of $x$, will it continue to be feasible if $x$ increases or decreases?

\begin{theorem} \label{lem:one_interval}
Assume a decomposition $\{\mathcal{Z}_1, \mathcal{Z}_2, \mathcal{Z}_3, \mathcal{Z}_4\}$ is feasible for $x=x_0\in [0, Q_1]$. If $x$ increases from $x_0$, then there exists a point denoted $x_1\in [x_0, Q_1]$ such that the decomposition is always feasible in interval $[x_0,x_1]$, and is always infeasible in interval $(x_1, Q_1]$. If $x$ decreases from $x_0$, then there exists a point denoted $x_2\in [0, x_0]$ such that the decomposition is always feasible in interval $[x_2,x_0]$, and is always infeasible in interval $[0, x_2)$.
\end{theorem}
\begin{IEEEproof}
Here we only prove the case when $x$ increases, as the case when $x$ decreases can be proved similarly.

For an $x$ (i.e., $Q_1^{\text{III}}$) value, the feasibility of decomposition $\{\mathcal{Z}_1, \mathcal{Z}_2, \mathcal{Z}_3, \mathcal{Z}_4\}$ is checked as follows: calculate $\lambda$ and $\zeta$ based on (\ref{e:lambda_final}) and (\ref{e:zeta_final}), calculate $d_{n,1}$ and $d_{n,2}$ based on (\ref{e:dn1_II_final}), (\ref{e:dn2_II_final}), and the calculated $\lambda$ and $\zeta$ values, and calculate $\mu_n$ and $\nu_n$ based on (\ref{e:dn1_by_dn2_II}), (\ref{e:dn2_by_dn1_II}), and the calculated $d_{n,1}$, $d_{n,2}$, $\lambda$ and $\zeta$ values. If all the values of $\lambda$, $\zeta$, $d_{n,1}$, $d_{n,2}$, $\mu_n$, and $\nu_n$ ($n\in \mathcal{N}_\text{II}$) are non-negative, then the decomposition $\{\mathcal{Z}_1, \mathcal{Z}_2, \mathcal{Z}_3, \mathcal{Z}_4\}$ is feasible; otherwise, it is infeasible.

Expressions (\ref{e:lambda_final}) and (\ref{e:zeta_final}) show that $\lambda$ and $\zeta$ are linear functions of $x$ (i.e., $Q_1^{\text{III}}$).

Expressions (\ref{e:dn1_II_final}) and (\ref{e:dn2_II_final}) show that $d_{n,1}$ and $d_{n,2}$ are linear functions of $\lambda$ and $\zeta$, and thus, are linear functions of $x$.

Expressions (\ref{e:dn1_by_dn2_II}) and (\ref{e:dn2_by_dn1_II}) show that $\mu_n$ and $\nu_n$ are linear functions of $\lambda$, $\zeta$, $d_{n,1}$, and $d_{n,2}$, and thus, are linear functions of $x$.

Overall, $\lambda$, $\zeta$, $d_{n,1}$, $d_{n,2}$, $\mu_n$, and $\nu_n$ ($n\in \mathcal{N}_\text{II}$) are all linear functions of $x$ (i.e., $Q_1^{\text{III}}$).

When $x=x_0$, as the decomposition $\{\mathcal{Z}_1, \mathcal{Z}_2, \mathcal{Z}_3, \mathcal{Z}_4\}$ is feasible, all the $\lambda$, $\zeta$, $d_{n,1}$, $d_{n,2}$, $\mu_n$, and $\nu_n$ ($n\in \mathcal{N}_\text{II}$) are non-negative. When $x$ increases from $x$, values of $\lambda$, $\zeta$, $d_{n,1}$, $d_{n,2}$, $\mu_n$, and $\nu_n$  linearly change accordingly. If at one point, say $x=x_1$, one of $\lambda$, $\zeta$, $d_{n,1}$, $d_{n,2}$, $\mu_n$, and $\nu_n$ decreases to value $0$, then we can see that for $x\in [x_0,x_1]$, the decomposition $\{\mathcal{Z}_1, \mathcal{Z}_2, \mathcal{Z}_3, \mathcal{Z}_4\}$ is always feasible, and for $x\in (x_1,Q_1]$, the decomposition is always infeasible.\footnote{As an extreme case, if $\lambda$, $\zeta$, $d_{n,1}$, $d_{n,2}$, $\mu_n$, and $\nu_n$ all keep non-negative when $x$ increases from $x_0$ to $Q_1$, then we have $x_1=Q_1$.}

This completes the proof.
\end{IEEEproof}
Remark: Theorem \ref{lem:one_interval} shows that if a decomposition $\{\mathcal{Z}_1, \mathcal{Z}_2, \mathcal{Z}_3, \mathcal{Z}_4\}$ is feasible for $x=x_0$, then there exists an interval of $x$ containing $x_0$ such that the decomposition is feasible inside the interval, and infeasible outside the interval.

%

Based on Lemma \ref{lem:V_increasing_concave}, Theorem \ref{lem:U_x_differential_mono}, and Theorem \ref{lem:one_interval}, we propose that seller 1 uses the following Algorithm 1 to select $x$ (i.e., $Q_1^{\text{III}}$).

\begin{algorithm}[H]
\caption{Searching procedure for $x$ (i.e., $Q_1^{\text{III}}$).}
\label{alg1_eq}
\begin{algorithmic}[1]
 \STATE {Set $x^*=0$, and $R^*=0$.}
 \STATE {Set $x^\dag=0$}
 \STATE {For $x=x^\dag$, find out all feasible Nash equilibria, and pick up the Nash equilibrium that maximizes the minimal unit-bandwidth revenue of the two sellers.}
 \STATE {Find (using bisection search) a point denoted $x_1$ such that the Nash equilibrium picked in Step 3 is feasible for $x\in [x^\dag, x_1]$, and infeasible for $x\in (x_1, Q_1]$.}
 \STATE {Set $x^\ddag = x_1 $.} 
 \STATE {The Nash equilibrium picked in Step 3 is feasible for $x\in [x^\dag, x^\ddag]$. For complexity reduction, approximately seller 1 considers that the Nash equilibrium picked in Step 3 is followed by both sellers when $x\in [x^\dag, x^\ddag]$. The revenue of seller 1 is $U(Q_1-x) + V(x)$. Here $U(Q_1-x)$ is the difference of two monotonically increasing functions of $x$ (from Theorem \ref{lem:U_x_differential_mono}), while $V(x)$ is an increasing function of $x$ (from Lemma \ref{lem:V_increasing_concave}). Thus, $U(Q_1-x) + V(x)$ can be viewed as the difference of two monotonically increasing functions of $x\in [x^\dag, x^\ddag]$. To maximize the difference of two monotonically increasing functions, a polyblock method can be used (please refer to \cite{RF_Joint, YJ} for details). Denote the optimal point as $\hat{x}$ and the corresponding revenue $U(Q_1-\hat{x}) + V(\hat{x})$ of seller 1 as $\hat{R}$.}
 \STATE{If $\hat{R}>R^*$, then set $x^*=\hat{x}$ and $R^*=\hat{R}$.}
 \STATE {If $x^\ddag=Q_1$, then terminate the algorithm, and output $x^*$.}
 \STATE {Set $x^\dag=x^\ddag$, and proceed to Step 3.}
 \end{algorithmic}
\end{algorithm}

\begin{table*}
\begin{center}
\caption{The number of decompositions without and with the aid of Theorem \ref{lem:complexity_reduction}.}\label{t:complexity}
\begin{tabular}{c|c|c|c|c|c|c|c}
\hline\hline $|\mathcal{N}_{\text{II}}|$ & 2 &4 &6 &8 &10 &15 &20\\
\hline Total number of decompositions & 16 & 256 & 4096 & $6.6\times 10^5$ & $1.0\times 10^6$ & $1.1\times 10^{9}$ & $1.1\times 10^{12}$\\
\hline Checked decompositions (with Theorem \ref{lem:complexity_reduction}) & 13 & 121 & 1093 & 9841 & $8.9 \times 10^5$ &  $2.2\times 10^7$ & $ 5.2 \times 10^9$ \\
\hline\hline
\end{tabular}
\end{center}
\end{table*}

In the algorithm, $x^*$ denotes the selection of seller 1 for $x$, and $R^*$ denotes the corresponding overall revenue of seller 1. For $x=x^\dag=0$, in Step 3 we first select the Nash equilibrium that maximizes the minimal unit-bandwidth revenue of the two sellers. In Steps 4 and 5, we find the interval of $x$, denoted $[x^\dag, x^\ddag]$, such that the selected Nash equilibrium is feasible inside the interval and infeasible when $x>x^\ddag$. We approximately consider that the Nash equilibrium is followed by both sellers for the interval $x\in [x^\dag, x^\ddag]$.\footnote{If $|\mathcal{N}_{\text{III}}| \leq 12$, then according to Theorem \ref{lem:equilibrium_unique}, the Nash equilibrium is the unique Nash equilibrium for $x\in [x^\dag, x^\ddag]$, and thus, is always followed by both sellers when $x\in [x^\dag, x^\ddag]$.} Then for $x\in [x^\dag, x^\ddag]$, seller 1's revenue $U(Q_1-x) + V(x)$ can be shown as the difference of two monotonically increasing functions of $x$. Existing methods in the literature (such as a polyblock algorithm) can be used to find the optimal value of $x\in [x^\dag, x^\ddag]$, denoted $\hat{x}$, such that the overall revenue of seller 1 is maximized. Then the $\hat{x}$ is a candidate for seller 1's selection of $x$. Then we set $x^\dag=x^\ddag$ in Step 9 and repeat the above procedure, and find other candidates for seller 1's selection of $x$. Among all the candidates, the one that has the maximal overall revenue of seller 1 is eventually selected by seller 1.

Overall, the strategies of the two sellers are as follows. In Epoch I, seller 2 derives its optimal strategy by solving Problem \ref{p:Epoch_I}. At the beginning of Epoch II, seller 1 uses Algorithm 1 to find the value of $x$, denoted $x^*$, Then in the non-cooperative game in Epoch II with $Q_1^{\text{III}}=x^*$, both sellers follow the Nash equilibrium that maximizes the minimal unit-bandwidth revenue of the two sellers. In Epoch III, seller 1 can derive its optimal strategy by solving Problem \ref{p:Epoch_III} with $Q_1^{\text{III}}=x^*$.

\section{Numerical Results} \label{s:numerical results}
\subsection{Verification of the Analysis}
We use numerical results by Matlab to verify the theoretical analysis in this paper. Since the spectrum leasing problem in Epoch I and Epoch III are both convex optimization problems, here we focus on Epoch II. At the beginning of Epoch II, seller 1 has spectrum bandwidth with amount $Q_1=100$, while seller 2 has available spectrum bandwidth with amount $Q_2^{\text{II}}=60$. We take $C_0=480$ and $C_1=1$.
The number of stages in Epoch III is $|\mathcal{N}_{\text{III}}|=3$.

\subsubsection{Effectiveness of Theorem \ref{lem:complexity_reduction}}
In this subsection, the effectiveness of Theorem \ref{lem:complexity_reduction} in complexity reduction is verified. Table \ref{t:complexity} lists the number of all possible decompositions and the number of decompositions that should be checked for feasibility with the aid of Theorem \ref{lem:complexity_reduction}. It is clear that using Theorem \ref{lem:complexity_reduction} can significantly reduce the number of decompositions that should be checked.

\subsubsection{Verification of Lemma \ref{lem:V_increasing_concave}}

\begin{figure}
\begin{center}
\includegraphics[angle=0,width=3in]{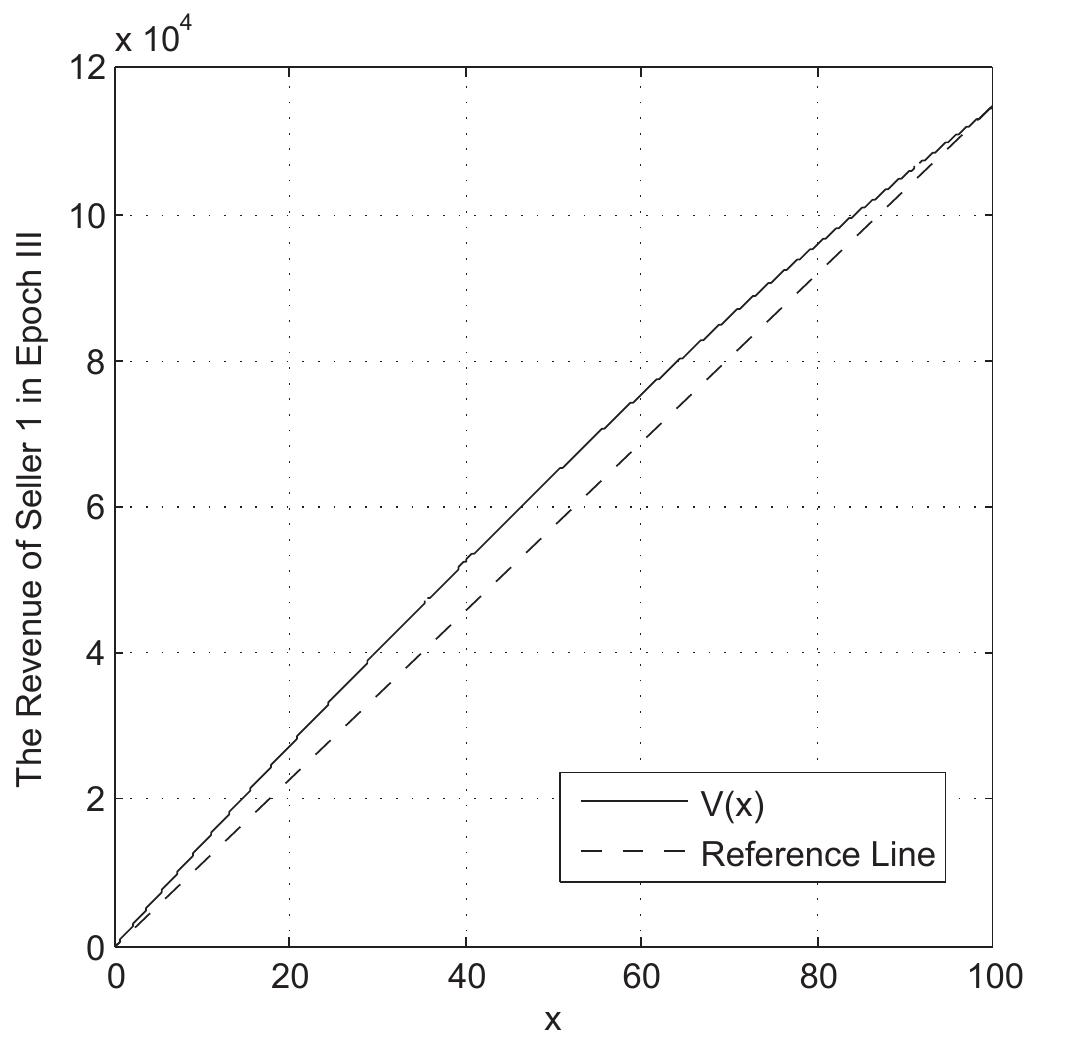}
\end{center}
\caption{$V(x)$ versus $x$ (i.e., $Q_1^{\text{III}}$).}
\label{fig_lem1}
\end{figure}

In this subsection, Lemma \ref{lem:V_increasing_concave} is verified. Fig.~\ref{fig_lem1} plots the function $V(x)$ (the revenue of seller 1 in Epoch III) as $x$ (i.e., $Q_1^{\text{III}}$) grows from 0 to 100.
From Fig.~\ref{fig_lem1}, it can be seen that the function $V(x)$ is an increasing and concave function with respect to $x$, which is consistent with Lemma \ref{lem:V_increasing_concave}. Note that the reference line in Fig. \ref{fig_lem1} is a straight line connecting points $(0, V(0))$ and $(100, V(100))$, which helps to observe the concavity of function $V(x)$.

\begin{figure}
\begin{center}
\includegraphics[angle=0,width=3in]{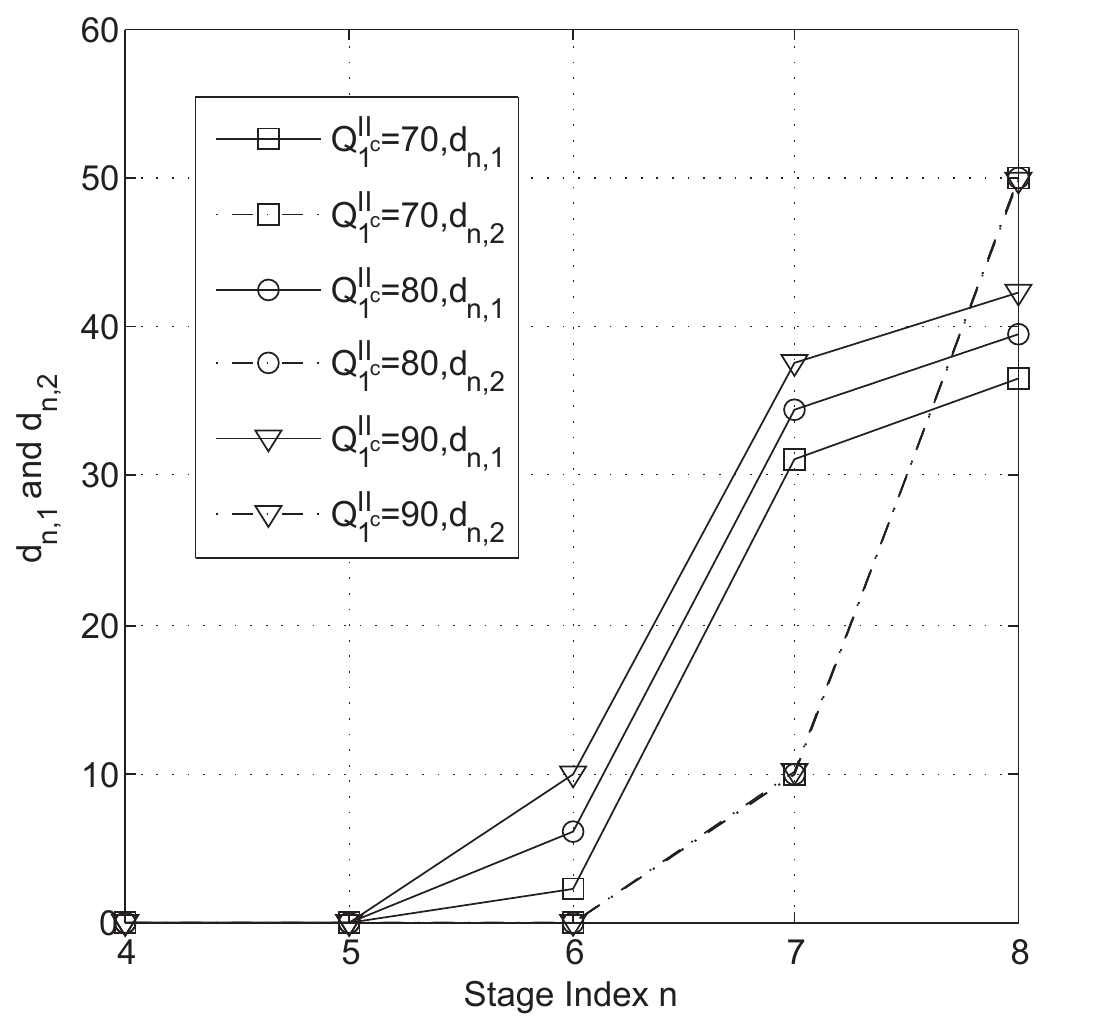}
\end{center}
\caption{$d_{n,1}$ and $d_{n,2}$ versus $n$ for $\mathcal{Z}_1=\{7,8\}$, $\mathcal{Z}_2=\{6\}$, $\mathcal{Z}_3=\emptyset$,  $\mathcal{Z}_4=\{4,5\}$, and $Q_1^{\text{II}_c}=70, 80, 90$.}
\label{fig_lem2}
\end{figure}

\begin{figure}
\begin{center}
\includegraphics[angle=0,width=3in]{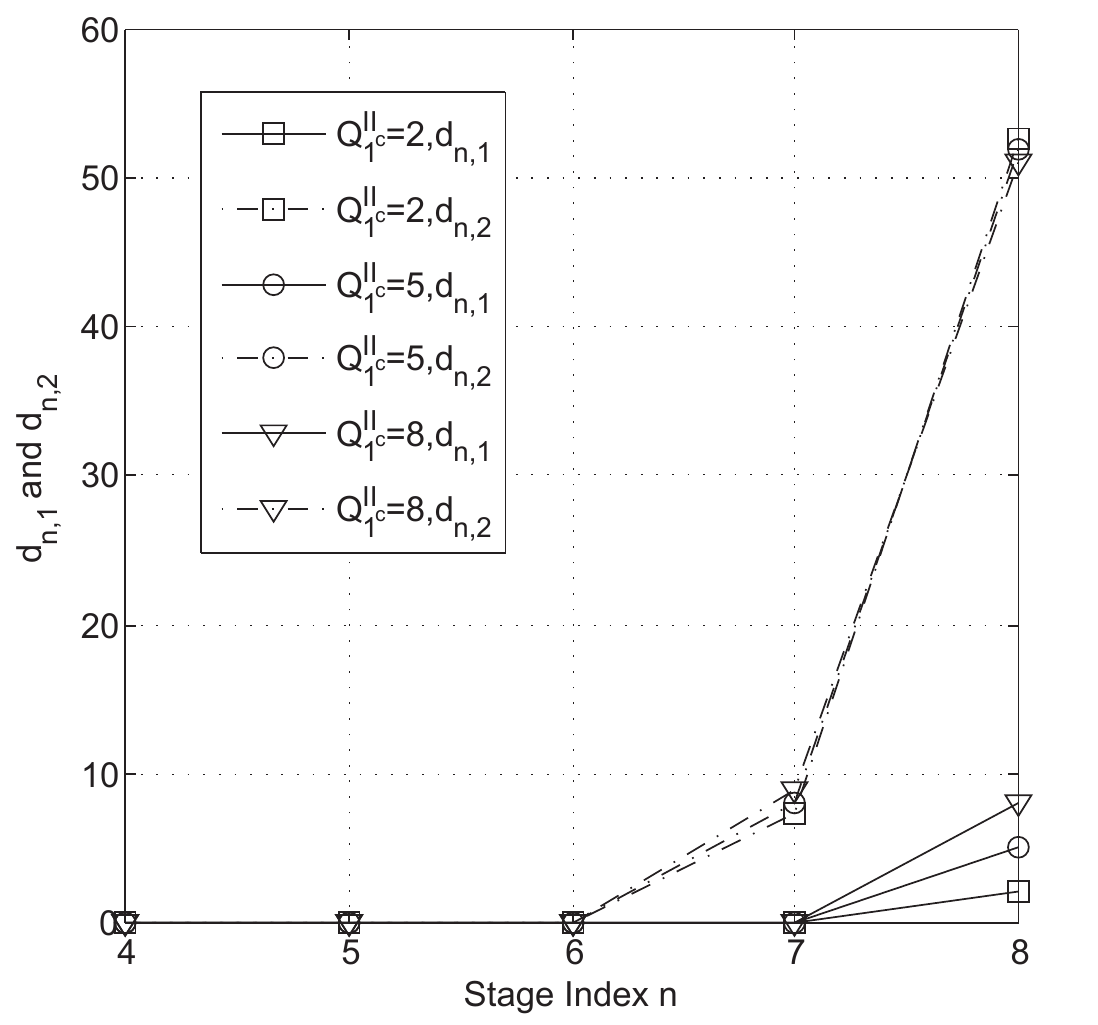}
\end{center}
\caption{$d_{n,1}$ and $d_{n,2}$ versus $n$ for $\mathcal{Z}_1=\{8\}$, $\mathcal{Z}_2=\emptyset$, $\mathcal{Z}_3=\{7\}$, $\mathcal{Z}_4=\{4,5,6\}$, and $Q_1^{\text{II}_c}=2, 5, 8$.}
\label{fig_lem3}
\end{figure}

\subsubsection{Verification of Lemma \ref{lem:d_1_mono} and Lemma \ref{lem:d_2_mono}}
In this subsection, Lemma \ref{lem:d_1_mono} and Lemma \ref{lem:d_2_mono} are verified. Consider $\mathcal{N}_\text{II}=\{4,5,6,7,8\}$.
Fig.~\ref{fig_lem2} plots $d_{n,1}$ and $d_{n,2}$ versus the stage index $n$ for a feasible decomposition in which $\mathcal{Z}_1=\{7,8\}$, $\mathcal{Z}_2=\{6\}$, $\mathcal{Z}_3=\emptyset$, and $\mathcal{Z}_4=\{4,5\}$ when $Q_1^{\text{II}_c}$ is set to be 70, 80, and 90.
Fig.~\ref{fig_lem3} plots $d_{n,1}$ and $d_{n,2}$ versus the stage index for a feasible decomposition in which $\mathcal{Z}_1=\{8\}$, $\mathcal{Z}_2=\emptyset$, $\mathcal{Z}_3=\{7\}$, and $\mathcal{Z}_4=\{4,5,6\}$ when $Q_1^{\text{II}_c}$ is set to be 2, 5, and 8.
From Fig. \ref{fig_lem2} and Fig. \ref{fig_lem3}, it can be seen that, when $Q_1^{\text{II}_c}$ changes, $d_{n,1}$ and $d_{n,2}$ vary in the same way as Lemma \ref{lem:d_1_mono} and Lemma \ref{lem:d_2_mono} describe.

\subsubsection{Verification of Theorem \ref{lem:U_x_differential_mono}}\label{s:ver_theorem3}
\begin{figure}
\begin{center}
\includegraphics[angle=0,width=3in]{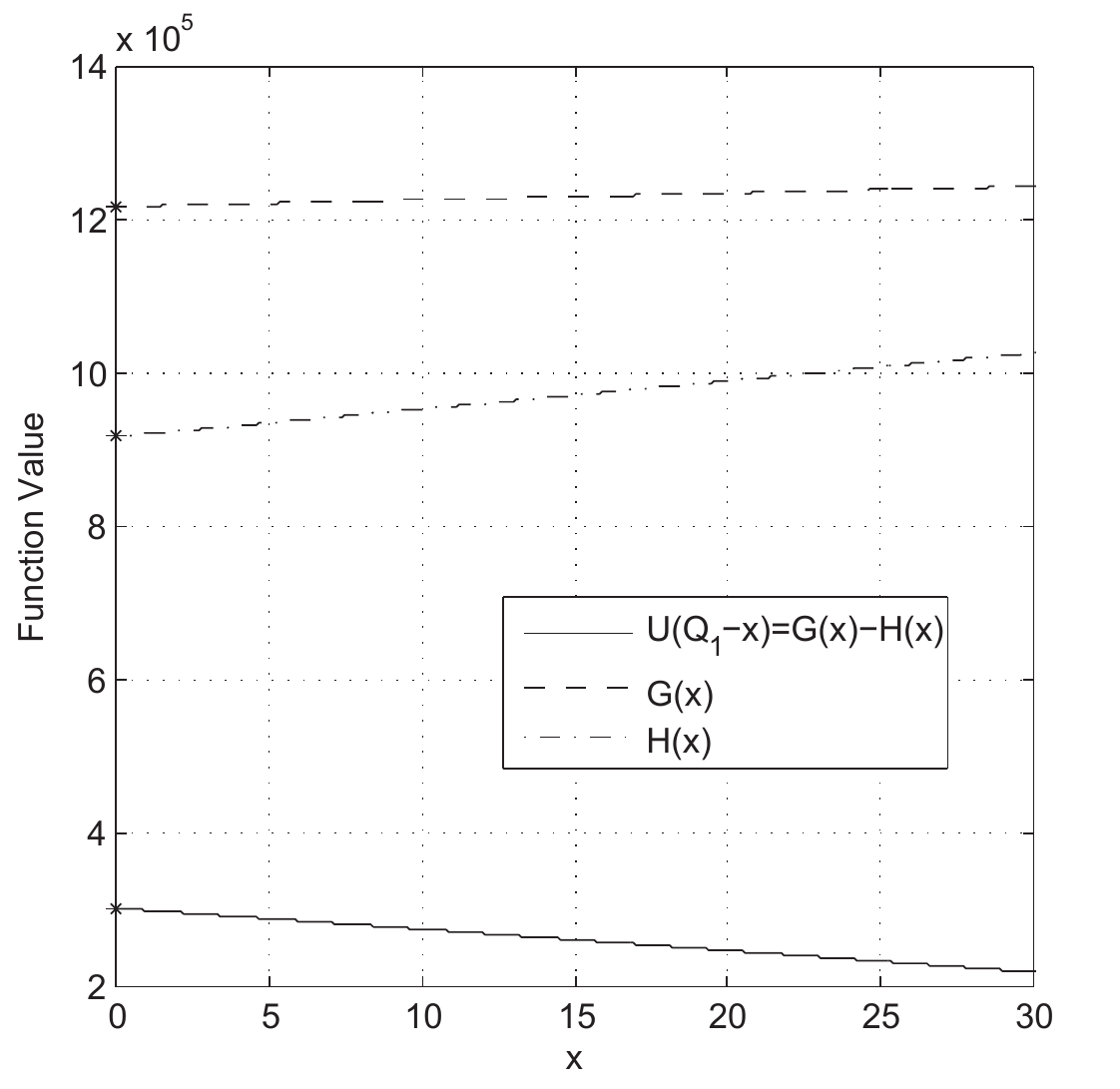}
\end{center}
\caption{Functions $U(Q_1- x)$, $G(x)$ and $H(x)$ ($x\in [0,30]$) for the first decomposition in Table \ref{tab_decomposition}.}
\label{fig_thm31}
\end{figure}

\begin{figure}
\begin{center}
\includegraphics[angle=0,width=3in]{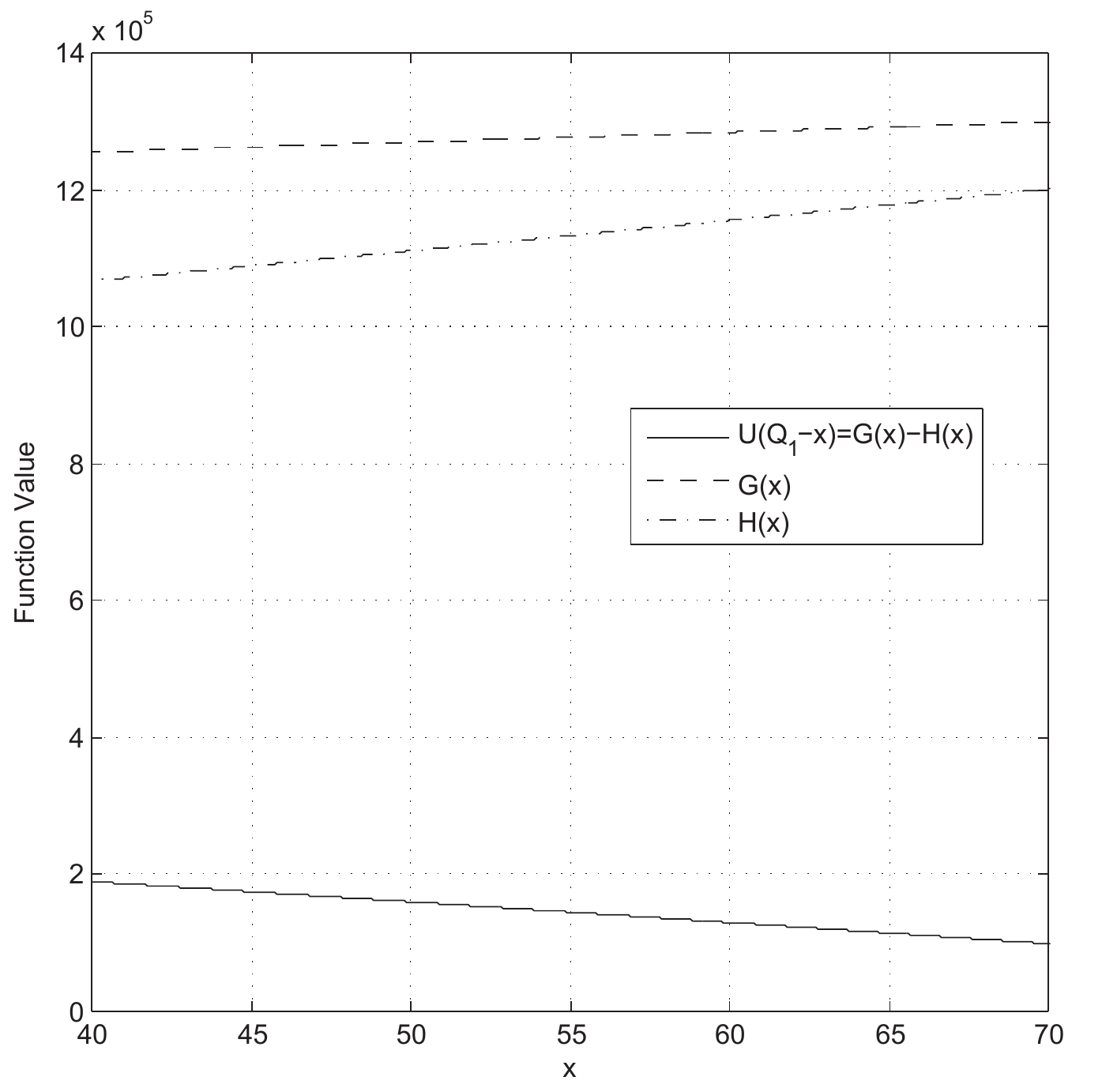}
\end{center}
\caption{Functions $U(Q_1- x)$, $G(x)$ and $H(x)$ ($x\in [40,70]$) for the second decomposition in Table \ref{tab_decomposition}.}
\label{fig_thm32}
\end{figure}

In this subsection, the characteristic of $U(Q_1-x)$ described in Theorem \ref{lem:U_x_differential_mono} is verified. Still consider $\mathcal{N}_\text{II}=\{4,5,6,7,8\}$. Two decompositions are investigated, which are listed in Table \ref{tab_decomposition}. Consider two intervals of $x$: $[0,30]$ and $[40,70]$, in which the two decompositions are feasible, respectively. Fig.~\ref{fig_thm31} and Fig.~\ref{fig_thm32} plot the function $U(Q_1- x)$ as well as $G(x)$ and $H(x)$ (from Theorem \ref{lem:U_x_differential_mono}) for the two decompositions over the two corresponding intervals, respectively. It can be seen that both the functions $G(x)$ and $H(x)$ are monotonically increasing for each decomposition in the corresponding  interval of $x$, which is consistent with Theorem \ref{lem:U_x_differential_mono}.

\begin{table}
\centering
\caption{The Decompositions used when verifying Theorem \ref{lem:U_x_differential_mono}}
\label{tab_decomposition}
\begin{tabular}{|c|c|l|c|c|}
\hline
\hline
 & $\mathcal{Z}_1$ & $\mathcal{Z}_2$ & $\mathcal{Z}_3$ & $\mathcal{Z}_4$ \\
\hline
 1st decomposition & $\{7,8\}$ & $\{6\}$ & $\emptyset$ & $\{4,5\}$ \\
\hline
 2nd decomposition & $\{7,8\}$ & $\emptyset$ & $\emptyset$ & $\{4,5,6\}$ \\
\hline
\hline
\end{tabular}
\end{table}

\begin{figure}
\begin{center}
\includegraphics[angle=0,width=3in]{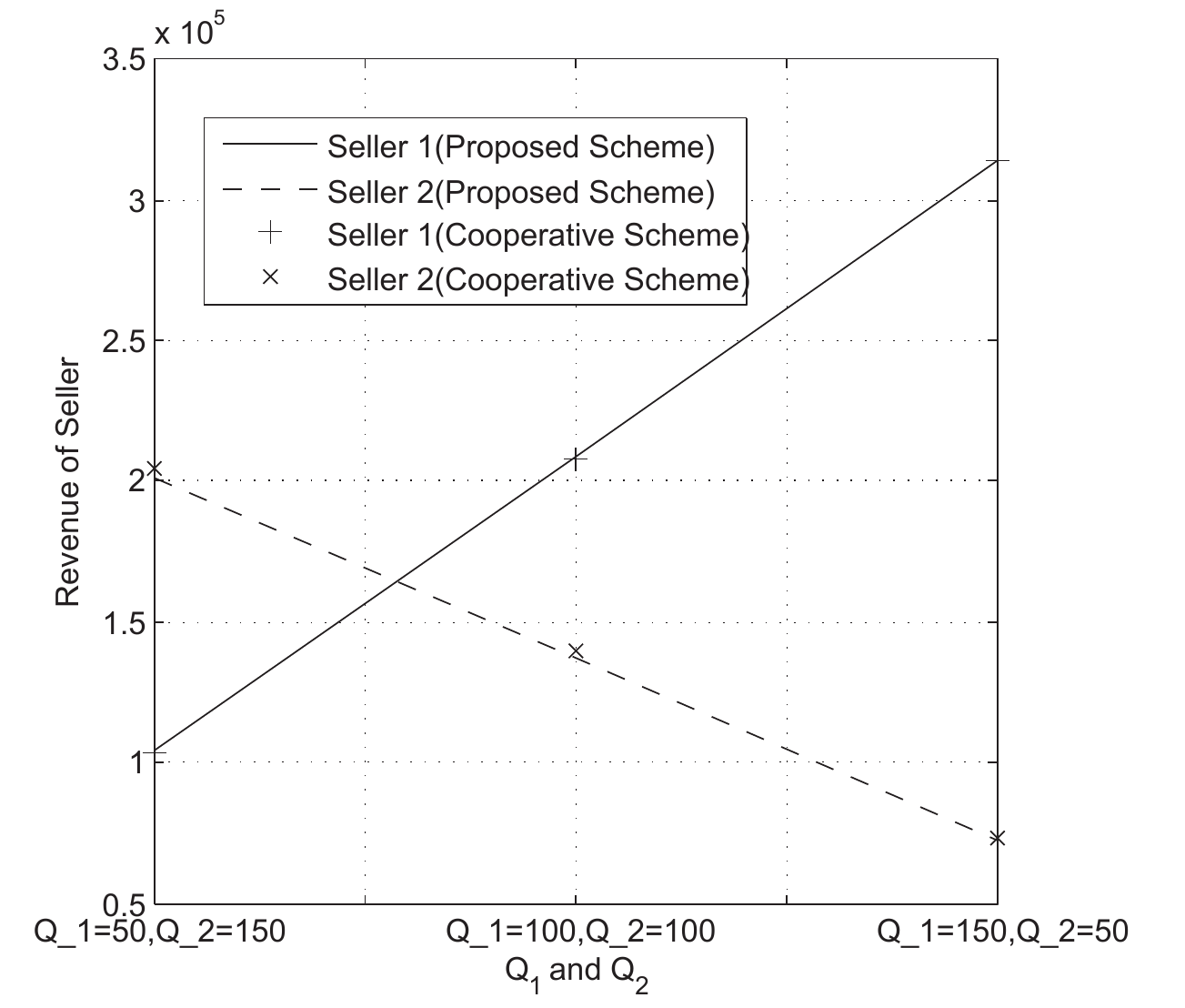}
\end{center}
\caption{The revenue of the two sellers in our proposed scheme and the cooperative scheme.}
\label{fig_comp}
\end{figure}

\subsection{Comparison with a cooperative scheme}
Now we compare with other schemes. As there is no research in the literature on dynamic pricing for more than one seller, we compare with a cooperative scheme. The difference of the cooperative scheme from our proposed scheme is as follows. When the two sellers know the existence of each other (i.e., at the beginning of Epoch II), the two sellers cooperate to jointly maximize the total revenue of them over Epoch II and III, by solving the optimization problem shown in (\ref{e:compared_al}).

\begin{equation}\label{e:compared_al}
\begin{array}{cll}
\mathop {\max} \limits_{\{d_{n, 1}|n \in \mathcal{N}_\text{II}\cup \mathcal{N}_\text{III}\},\{d_{n, 2}|n \in \mathcal{N}_\text{II}\}} & \sum\limits_{n \in \mathcal{N}_\text{II}} \left(C_0-C_1 (d_{n,1}+d_{n,2})\right)  d_{n,1} n + \sum\limits_{n \in \mathcal{N}_\text{III}}  \left(C_0-C_1 d_{n,1}\right)  d_{n,1} n \\&+\sum\limits_{n \in \mathcal{N}_\text{II}}  \left(C_0-C_1 (d_{n,1}+d_{n,2})\right)  d_{n,2} (n- |\mathcal{N}_{\text{III}}|) \\
\text{s.t.} & \sum \limits_{n \in \mathcal{N}_\text{II}\cup \mathcal{N}_\text{III}} d_{n,1} \leq Q_{1} \\
& \sum \limits_{n \in \mathcal{N}_\text{II}} d_{n,2} \leq Q_{2}^\text{II}\\
& d_{n,1} \geq 0, \forall n \in \mathcal{N}_\text{II}\cup \mathcal{N}_\text{III} \\
& d_{n,2} \geq 0, \forall n \in \mathcal{N}_\text{II}.
\end{array}
\end{equation}


For performance comparison, the simulation is set up as follows. Since the cooperative scheme and our proposed scheme perform the same in Epoch I, we set $\cal{N}_\text{I}=\emptyset$. And $\mathcal{N}_\text{II}=\{6,5,4,3\}$, $\mathcal{N}_\text{III}=\{2,1\}$. We fix the sum of $Q_1$ and $Q_2$ to be 200, and consider three configurations of $(Q_1,Q_2)$: $(50,150), (100,100)$, and $(150,50)$. Fig.~\ref{fig_comp} shows the achieved revenue of the two sellers in our proposed scheme and the cooperative scheme. It can be seen that each seller's revenue in our proposed non-cooperative scheme is very close to that in the cooperative scheme, thus verifying the efficiency of our proposed scheme.

\section{Conclusions} \label{s:conclusion}
In this paper, we investigate spectrum leasing with two sellers. In Epoch II, the two sellers both have spectrum to lease, and competition between the two sellers exists. Thus, the spectrum leasing in Epoch II is formulated as a non-cooperative game. Nash equilibria of the game are derived in closed form by jointly solving two optimization problems. By analyzing the choices of seller 1, solutions of the two sellers in the spectrum leasing are developed. The analysis and solutions in this work should help design oligopoly spectrum leasing strategies in future cognitive radio networks.

\end{document}